# A framework for automated anomaly detection in high frequency water-quality data from *in situ* sensors

**Authors**:
Catherine Leigh[1,2,3], Omar Alsibai[1,2], Rob J. Hyndman[1,4], Sevvandi Kandanaarachchi[1,4], Olivia C. King[5], James M. McGree[1,3], Catherine Neelamraju[5], Jennifer Strauss[5], Priyanga Dilini Talagala[1,4], Ryan D. R. Turner[5], Kerrie Mengersen[1,3], Erin E. Peterson[1,2,3]

**Affiliations:**
[1] ARC Centre of Excellence for Mathematical & Statistical Frontiers (ACEMS), Australia
[2] Institute for Future Environments, Queensland University of Technology, Brisbane, Queensland, Australia
[3] School of Mathematical Sciences, Science and Engineering Faculty, Queensland University of Technology, Brisbane, Queensland, Australia
[4] Department of Econometrics and Business Statistics, Monash University, Clayton, Victoria, Australia.
[5] Water Quality and Investigations, Department of Environment and Science, Dutton Park, Queensland, Australia.

**Corresponding author:** catherine.leigh@qut.edu.au

**Highlights**

- High frequency water-quality data requires automated anomaly detection (AD)
- Rule-based methods detected all missing, out-of-range and impossible values
- Regression and feature-based methods detected sudden spikes and level shifts well
- High false negative rates were associated with other types of anomalies, e.g. drift
- Our transferable framework selects and compares AD methods for end-user needs

# Abstract

Monitoring the water quality of rivers is increasingly conducted using automated *in situ* sensors, enabling timelier identification of unexpected values or trends. However, the data are confounded by anomalies caused by technical issues, for which the volume and velocity of data preclude manual detection. We present a framework for automated anomaly detection in high-frequency water-quality data from *in situ* sensors, using turbidity, conductivity and river level data collected from rivers flowing into the Great Barrier Reef. After identifying end-user needs and defining anomalies, we ranked anomaly importance and selected suitable detection methods. High priority anomalies included sudden isolated spikes and level shifts, most of which were classified correctly by regression-based methods such as autoregressive integrated moving average models. However, incorporation of multiple water-quality variables as covariates reduced performance due to complex relationships among variables. Classifications of drift and periods of anomalously low or high variability were more often correct when we applied mitigation, which replaces anomalous measurements with forecasts for further forecasting, but this inflated false positive rates. Feature-based methods also performed well on high priority anomalies and were similarly less proficient at detecting lower priority anomalies, resulting in high false negative rates. Unlike regression-based methods, however, all feature-based methods produced low false positive rates and have the benefit of not requiring training or optimization. Rule-based methods successfully detected a subset of lower priority anomalies, specifically impossible values and missing observations. We therefore suggest that a combination of methods will provide optimal performance in terms of correct anomaly detection, whilst minimizing false detection rates. Furthermore, our framework emphasizes the importance of communication between end-users and anomaly detection developers for optimal outcomes with respect to both detection performance and end-user application. To this end, our framework has high transferability to other types of high frequency time-series data and anomaly detection applications.

**Abbreviations**
AD, anomaly detection
ADAM, anomaly detection and mitigation
ARIMA, autoregressive integrated moving average
FN, false negative
FP, false positive
PI, prediction interval
PR, Pioneer River
RegARIMA, multivariate regression with ARIMA errors
SC, Sandy Creek
TN, true negative
TP, true positive

# 1. Introduction

Clean water is a United Nations Sustainable Development Goal as well as a major concern in many developed areas. Monitoring the quality of water in the world's rivers relies predominantly on manual collection of water-quality samples at low frequencies (e.g. monthly). These discrete samples are analysed in laboratories to provide estimates of the concentrations of ecologically important constituents such as sediments and nutrients. This requires considerable time and money, and the resulting data are typically sparse in space and time. Fortunately, other properties of water, such as turbidity and conductivity, can be measured semi-continuously by readily available, low-cost, automated *in situ* sensors, and show promise as surrogates of sediment and nutrient concentrations in rivers (Jones et al., 2011; Slaets et al., 2014). Nevertheless, technical issues in sensor monitoring equipment can occur, for example, when battery power is low or sensors drift over time due to biofouling of the probes, or due to errors in calibration. These issues can lead to errors in water-quality measurements, which we define herein as anomalies. Such anomalies can be important to detect because they can confound the assessment or identification of true changes in water quality.

Notwithstanding technical errors, another issue that mitigates the potential advantages of using *in situ* sensor data is the production of high-frequency water-quality data in near-real time (i.e. data streaming). This high velocity, high volume data creates problems for quality control and assurance, given that manual checking, labelling and correction of each observation is unfeasible (Hill and Minsker, 2010; Horsburgh et al., 2015). We therefore need to develop robust methods for automatic anomaly detection (AD) before water-quality data from *in situ* sensors can be used with confidence for water-quality visualization, analysis and reporting.

Here, our objective was to develop a ten-step AD framework to implement and compare a suite of AD methods for high-frequency water-quality data measured by *in situ* sensors (Figure 1). We demonstrate this framework using a real-world case study where turbidity, conductivity and river level data were measured by automated *in situ* sensors in rivers flowing into the Great Barrier Reef lagoon of northeast Australia. Anomalies were defined as any water-quality observations that were affected by technical errors in the sensor equipment; in other words, not due to real ecological patterns and processes occurring within the rivers and watersheds being monitored. We focussed on AD in turbidity and conductivity data because these properties of river water are typically more stable through time than other properties such as dissolved oxygen and temperature, which fluctuate daily as well as seasonally. Turbidity and conductivity were also the target water-quality variables for the end-user in our case study, described in Sections 2.1-2. We present this framework below and discuss the results of AD for high-frequency water-quality data from automated *in situ* sensors, with a view to providing insight on broader applications and future directions.

# 2. Methods

We describe below the method components of the AD framework (Steps 1 to 8; Figure 1) from identifying end-user needs and anomaly types and priorities through to selecting suitable analytical methods of AD.

## 2.1 Identify end-user needs and goals (Step 1)

Identifying the needs and goals of the end-user is a key step in the AD framework because this helps determine which types of anomalies will be most important to detect and thus the most suitable AD methods (Figure 1, Table 1). For this case study, several discussions were held between the end-user (an environmental agency concerned with water quality monitoring and management), statisticians and freshwater scientists prior to commencing analysis. The foremost, short-term need of the environmental agency was to produce 'smart' graphical outputs of the streaming water-quality data from *in situ* sensors for visualization in near-real time (Table 1). Visualization of streaming water-quality data helps to engender confidence in those data, and this means that potentially anomalous water-quality observations need to be identified and labelled as such, in near-real time. The longer-term goals of the end-user, beyond the specific scope of this case study, were to provide complete quality control and assurance of the data; flagging potential anomalies in near-real time, potentially with automated correction, and ultimately to use the corrected data to estimate sediment and nutrient concentrations in rivers in near-real time. Resultant data can then be used for accurate load estimation across multiple temporal scales. For other end-users, for example, the public, priority goals may include on-line and real-time warning of water quality breaches associated with real events (rather than technical anomalies). This may have serious economic and public health consequences in practice, affecting commercial operations (e.g. fisheries and aquaculture) and recreational sites (e.g. Rabinovici et al. 2004).

## 2.2 Identify data characteristics (Step 2)

Temporal characteristics of the time series data on which AD is performed play a role in determining the types of methods most suitable to use (Figure 1). Here, we used water-quality data from *in situ* sensors deployed in rivers of tropical northeast Australia that flow into the Great Barrier Reef lagoon. The rivers of interest are located in the Mackay Whitsunday region, east of the Great Dividing Range in Queensland, Australia. This region is characterized by a highly seasonal climate, experiencing higher rainfall and air temperatures in the 'wet' season (typically occurring between December and April and associated with event flows in rivers) and lower rainfall and air temperatures in the 'dry' season (associated with low to zero flows in rivers; Brodie, 2004). These characteristics affect the patterns of water quality in these rivers through time.

We focused on two rivers in the study region: Pioneer River and Sandy Creek. The upper reaches of Pioneer River flow predominantly through National or State Parks, with its middle and lower reaches flowing through land dominated by sugarcane farming. Sandy Creek is a low-lying coastal-plain stream, south of the Pioneer River, with a similar land-use and land-cover profile to that of the lower

Pioneer River. Two study sites, one on Pioneer River and one on Sandy Creek (PR and SC, respectively), are in freshwater reaches and their monitored catchment areas are 1466 $km^2$ and 326 $km^2$, respectively.

Automated water-quality sensors (YSI EXO2 Sondes with YSI Smart Sensors attached) have been installed at each site, housed in flow cells in water-quality monitoring stations on riverbanks. At each site, a pumping system is used to transport water regularly from the river up to the flow cell, approximately every hour or hour and a half, for the sensors to measure and record turbidity (NTU) and electrical conductivity at 25 °C (conductivity; µS/cm). All equipment undergo regular maintenance and calibration, with sensors calibrated and equipment checked approximately every 6 weeks following manufacturer guidelines. Sensors are equipped with wipers to minimise biofouling. Pre-wet season maintenance, e.g. cleaning samplers and drainage lines from the flow cell, is performed annually.

Turbidity is an optical property of water that reflects its level of clarity, which declines as the concentrations of abiotic and biotic suspended particles that absorb and scatter light increase. Turbidity thus tends to increase rapidly during runoff events when waters contain high concentrations of suspended particles eroded from the land and upstream river channels. When waters concentrate during times of low or zero flow, turbidity may increase gradually through time. Similarly, conductivity, which reflects the concentration of ions including bioavailable nutrients such as nitrate and phosphate in the water, also tends to increase during periods of low and zero flow, but can decrease rapidly with new inputs of fresh water. Measurements of turbidity and conductivity may be taken more frequently during event flows to capture the increased range of values observed during such times; however, the relationships among river level, turbidity and conductivity are complex (Figure S1). River level (i.e. height in meters from the riverbed to the water surface; level, m) is recorded by sensors at each site every 10 minutes; we linearly interpolated these data to provide time-matched observations of level for each turbidity and conductivity observation. Although we did not perform anomaly detection on the river level data, we examined its relationship with the turbidity and conductivity data to provide insight into the water-quality dynamics occurring at both study sites (Figure S1). The time series data from the *in situ* sensors were available from 12 March 2017 to 12 March 2018 at both sites, totalling 6280 and 5402 observations at PR and SC, respectively (Figures S2-S3).

## 2.3 Define anomalies and their types (Step 3)

A clear definition of what constitutes an anomaly, relevant to the data and end-user requirements, is needed prior to commencing detection (Figure 1). Several definitions of anomalies exist, each differing in specificity. In general, they are considered to (i) differ from the norm with respect to their features, and (ii) be rare in comparison with the non-anomalous observations in a dataset (Goldstein and Uchida, 2016). As mentioned, we defined an anomaly here as any water-quality datum or set of data that was due to a technical error in the *in situ* sensor equipment. For example, a real datum might include a rare, high-magnitude value of turbidity associated with heavy, erosive local rainfall and an ensuing high-flow event, whereas an anomaly might be a similarly high datum but one that is beyond the range of detection by the sensor.

Once ‘anomaly’ is defined, the different types of anomalies likely to be present in the time series data of interest must be defined and identified. We defined the different types of anomalies likely to occur in the water-quality data, in consultation with the end-user in this study, as: sudden spikes (large A, small J), low variability including persistent values (B), constant offsets (C), sudden shifts (D), high variability (E), impossible values (F), out-of-sensor-range values (G), drift (H), clusters of spikes (I), missing values (K) and other, untrustworthy (L; not described by types A-K) (Table 1, Figure 2). Some of these types have been described elsewhere for high frequency water-quality data, using the same or different terminology (e.g. Horsburgh et al., 2015), while other types were identified as more relevant to the specific characteristics of the data we were analysing (e.g. periods of anomalously high variation; Table 1). Other terms, such as local and global anomalies, have been used to describe anomalies in other contexts. We do not use these other terms here, chiefly because they do not adequately differentiate between the relevant types of anomalies we defined. For example, local anomalies, as defined by Goldstein and Uchida (2016), are only anomalous when compared with their immediate neighbourhood of data values. These may include large or small sudden spikes, values that are anomalously different in magnitude to that of data at neighbouring time steps. Global anomalies, on the other hand, are anomalously different to the majority of other data points, regardless of time (Goldstein and Uchida, 2016). Contextual anomalies describe data that appear anomalous only when context (e.g. season) is taken into account, otherwise appearing 'normal' (Goldstein and Uchida, 2016). For example, a high value of river level may be non-anomalous in the wet season, but could be anomalous within the context of the dry season. Contextual anomalies may, for example, include some anomalies identified here as type L (other, untrustworthy data). Types B, E, H and I anomalies may be referred to elsewhere as collective anomalies, i.e. collections of anomalous data points (Chandola et al., 2009). We additionally labelled the first observation after an extended period of missing data (i.e. no observations for more than 180 minutes) to identify it as an anomaly type K (see also Section 2.5.1).

We grouped the anomaly types into three general classes (Table 1). Class 1 included anomalies described by a sudden change in value from the previous observation (types A, D, I, and J). Class 2 included those anomaly types that should be detectable by simple, hard-coded classification rules, such as measurements outside the detectable range of the sensor (types F, G and K), whereas Class 3 anomalies may require user intervention *post hoc* (i.e. after data collection rather than in real time) to confirm observations as anomalous or otherwise in combination with automated detection (types B, C, E, H and L). Finally, we noted the times at which sensor maintenance activities such as probe swapping for calibration purposes were performed, to flag when anomalies might be likely to occur and provide causal insight about anomaly generation (Figures S2-S3).

We visually examined the water-quality time series data in consultation with the end-user. The potential anomalies in each time series at each site were identified and labelled along with their types based on expert knowledge of riverine water-quality dynamics and the particular sites and watersheds of interest. The labelled anomalies were used in steps 8-9 to implement AD and assess its performance.

## 2.4 Rank anomaly types by importance (Step 4)

The importance ranking for anomaly types is based on the potential impact each type may have if it were to go undetected, with respect to end-user needs and goals. This ensures that the end-user can effectively assess the ability of the AD methods to identify the most important anomalies. For example, one method may detect the same amount of anomalies as another; whilst the first method identifies anomalous high-magnitude values in a turbidity time series, the second method instead identifies minimally negative (impossible) values during periods of low turbidity only. If the end-user deems the former type of anomaly as more important to detect, then this would affect the evaluation of which AD method performs best and is most suitable. The rationale for the ranking might be that high-magnitude anomalies falsely indicate a breach of water-quality guidelines, whereas the change in turbidity caused by the negative readings may be negligible in the context of the period in which they occurred.

Here, we liaised with the end-user (in this case, an environmental agency concerned with water management and monitoring, see Section 2.1) to rank the importance of the different anomaly types identified as per Section 2.3 (Table 1). Their first priority was to identify large sudden spikes (Type A, Class 1) given that the short-term aim of anomaly detection was time series visualization. Sudden shifts (Type D, Class 1), calibration offsets (Type C, Class 3) and changes in variance (Types B and E, Class 3) were also deemed important, ranking second to fourth in priority, with types C and D both ranked third in place (Table 1).

## 2.5 Select suitable methods of anomaly detection (Step 5)

As outlined in Step 2 (Section 2.2), characteristics of the data on which AD is performed play a role in determining the most suitable AD methods, taking the end-user needs into account. Time series data are typically nonstationary, such that statistical parameters of the data (e.g. the mean and standard deviation) change with time. Furthermore, the production of high-frequency water-quality data from *in situ* sensors in near-real time creates 'big data', i.e. high-volume, high-velocity and high-variety information (Gandomi and Haider, 2015). This may be problematic for certain AD methods such as those developed for or typically applied to relatively small batches of pre-collected (historical) data (Liu et al., 2015).

We reviewed and compared the different AD methods used for water quality and time series data as described in the literature to identify those that are, or could be made, suitable for analysing near-real time and nonstationary data streams (Table S1). This included automated classification rules as well as several regression and feature-space based methods. Many of these methods are well documented and freely available software is available to implement them. Thus, they serve as suitable benchmarks for new anomaly detection methods that may be developed in the future. We chose to implement a suite of these methods because different algorithms are likely to detect certain types of anomalies (e.g. priority anomalies like large sudden spikes; Table 1) better than others.

Although we also considered physical-process based models for AD in water-quality time series (e.g. Moatar et al., 2001; Table S1), we did not explore them further here. Variation in water-quality patterns through time in rivers, and the multiple interactions within and between water-quality variables and the broader environment creates complexities and uncertainties that can make development of such models challenging and limit their transferability (e.g. Cox, 2003), particularly in the context of streaming data. Likewise, we did not explore dynamic Bayesian networks or hidden Markov models (Table S1). While both methods show potential in the context of streaming time series data (Hill et al., 2009; Li et al., 2017), their application in this context is relatively new with limited existing software for implementation using water-quality data.

### 2.5.1 Automated classification rules

Perhaps the simplest way to detect and classify anomalies such as impossible, out-of-sensor-range and missing values (Class 2: type F, G and K, respectively) is to develop rules that can be automated and applied to the streaming data in near-real time in combination with data-driven approaches such as regression and feature-based AD (see Section 2.5.2-3). For instance, negative values are impossible for turbidity and a simple rule (e.g. a 'range test'; Fiebrich et al., 2010) could therefore be set to classify any negative turbidity observation as an anomaly. Here, we implemented 'if-then' statements to detect and classify Class 2 anomalies. The first statement classified type K anomalies, using an end-user defined period as the maximum allowable time between two consecutive observations before the second observation is classed as a K, indicating that it occurred after a period of missing observations. Here we defined the maximum allowable threshold as 180 minutes (3 hours); however, this definition will vary according to end-user needs and the frequency of the *in situ* sensor data. We next identified type F anomalies (i.e. impossible values); if a turbidity or conductivity observation was negative, then it was classed as an anomaly. Furthermore, if any turbidity or conductivity observation was zero, then it was likewise classed as an anomaly because completely clear river water containing zero positive or negative ions is unrealistic. The if-then statements used to detect type G anomalies were based on range tests defined by sensor specifications for each water-quality variable.

### 2.5.2 Regression-based methods

The regression-based approach to AD in time series has several advantages, including (for some methods) the ability to deal with nonstationarity and provide near-real time support (Table S1). Furthermore, information from single or multiple water-quality variables as well as previous measurements can be taken into account, which makes these methods useful in the context of AD for streaming water-quality data. Most regression-based methods used for AD are semi-supervised (Table S1); the models are trained to learn the ‘normal’ (i.e. non-anomalous, typical) behaviour in a time series and, theoretically, should then detect any non-normal (i.e. anomalous) behaviour, regardless of the underlying cause.

To perform AD, the regression-based methods are used to generate a prediction, or forecast, with an associated measure of uncertainty at the next time point. The prediction intervals should ideally account for uncertainty associated with the model, model parameter values and the behaviour of future data, although in practice often only the model uncertainty is taken into account (Hyndman and Athanasopoulos, 2018). If the one-step-ahead observation does not fall within the prediction interval, it is classified as an anomaly.

The general form for regression-based methods can be written as:

$$x_t = \beta' Z_t + \eta_t$$
$$\eta_t = ARIMA(p, d, q)$$

where $x_t$ is the observation at time *t*, $\beta'$ is a vector of regression coefficients, and $Z_t$ is a vector of covariates. Thus, the errors from the regression model may be serially correlated, and we model this correlation structure using an ARIMA model. ARIMA models are discussed further below, and in detail in Hyndman and Athanasopoulos (2018), and can be thought of as a nonlinear regression against past observations. We assume the ARIMA model errors are uncorrelated in time, and normally distributed with zero mean, and we denote this by $\varepsilon_t \sim N(0, \sigma)$.

We let $\tilde{x}_{t+1}$ denote the one-step forecast of $x_{t+1}$ made at time *t*. To compute these forecasts, we add $\beta' Z_{t+1}$ to the forecasts from the ARIMA model.

After forecasting, observations are classified as anomalies, or not, based on the specified prediction interval. There is no training involved in this step. Here, we constructed a 100(1-α)% prediction interval (*PI*) for the one-step-ahead prediction (the forecast observation at time *t* + 1) as:

$$PI_{t+1} = \tilde{x}_{t+1} \pm t_{\alpha/2, T-k} \times s$$

where *T* is the size of the training dataset, *k* is the number of parameters in the model, $t_{\alpha/2,T-k}$ is the $\alpha/2$ quantile of a *t*-distribution with $T$ - *k* degrees of freedom, and $s$ is the square root of the mean of the squared ARIMA residuals in the training dataset.

The *PI* defines the range of ‘normal’ (i.e. non-anomalous) one-step-ahead predictions. The choice of significance level therefore affects the number of false positives produced. There were relatively few labelled anomalies in our time series data, especially for certain water-quality variables and anomaly types (Table 2). We therefore used a 99% prediction interval ($\alpha$ = 0.01) to effectively limit the probability of false positives to 1%.

We implemented the following set of regression-based models, based on the general form, to detect anomalies in the turbidity and conductivity time series: (i) naïve prediction, (ii) linear autoregression, (iii) ARIMA models, and (iv) multivariate linear regression with ARIMA errors (RegARIMA).

Naïve prediction is a regression-based method that uses the most recent observation as the one-step-ahead forecast:

$$\tilde{x}_{t+1} = x_t$$

In the notation of our general model, $\beta = Z_t = 0$ and $\eta_t = ARIMA(0,1,0)$. The method assumes the one-step-ahead forecast depends only on the previous observation, therefore the only parameter to estimate is $s$, the square root of the mean squared residuals, where the residuals in this case are the differences between consecutive observations. Naïve prediction therefore does not require stationarity in the mean of the time series (Table S1).

Linear autoregression (Box and Jenkins, 1970) differs from naïve prediction because it gives a forecast that is a linear combination of the $p$ previous observations, rather than just the single previous observation:

$$\tilde{x}_{t+1} = c + \sum_{i=1}^{p} \phi_i \, x_{t-i}$$

where the constant *c* and the set $\{\phi_1, \phi_2, \ldots, \phi_p\}$ are model parameters estimated from the training data. In the notation of our general model, $c = \beta,\ Z_t = 1$ and $\eta_t = ARIMA(p, 0{,}0)$. We used the partial autocorrelation function (PACF) to select the optimal value of *p* for the linear autoregression models (Tsay, 1989).

The ARIMA(*p*,*d*,*q*) model introduced by Box and Jenkins (1970) is more generalized than naïve prediction or linear autoregression models and includes autoregressive (*p*), differencing (*d*) and moving average (*q*) components (i.e. the succession of averages calculated from successive segments of the time series). Here, *p* determines the number of previous observations (time lags) in the autoregressive model, *d* determines the number of differences between observations to use, and *q* determines the number of moving average terms (see also Hyndman and Athanasopolous 2018). ARIMA models can handle stationary as well as nonstationary time series by adding a differencing component, i.e. using $d > 0$. To decide on the optimal value of the *p*, *d* and *q* ARIMA components, we used an automated procedure, based on the Akaike information criterion (AIC; Akaike 1974); minimizing the AIC is asymptotically equivalent to using cross-validation (Hyndman and Athanasopoulos 2018).

Finally, RegARIMA models, also known as dynamic regression models, are a combination of ARIMA time series modelling and multivariate regression (Hyndman and Athanasopoulos, 2018), where multivariate regression uses information from multiple water-quality variables for forecasting the one-step-ahead prediction:

$$\tilde{x}_{t+1} = \beta_0 + \sum_{i=1}^{k} \beta_i z_{i,t+1} + \tilde{\eta}_{t+1}$$

where $z_{i,t+1}$ represents variable *i* from the set of variables *{1,...,k}* at some time $t + 1$. In this way, information from multiple variables are included in the model in addition to information provided by previous observations. Here we included turbidity and river level, or conductivity and river level, in the multiple regression component of the ARIMA model to forecast the one-step-ahead conductivity, or turbidity observations, respectively, using the AIC to select the best *p*, *d* and *q* parameters as per ARIMA above.

For all of the above methods we investigated assumptions of the models by conducting Box-Ljung portmanteau tests to assess stationarity in the mean (Ljung and Box, 1978) and producing diagnostics plots to visually assess stationarity in variance.

One additional approach to AD within the regression-based suite of methods, applied to water-quality time series by Hill and Minsker (2010), uses anomaly mitigation (i.e. correction) during forecasting and classification. Essentially this anomaly detection and mitigation (ADAM) approach uses forecasts instead of actual observations, when detected as anomalous, to forecast the subsequent one-step-ahead observation. ADAM therefore has the potential to change the regression forecasting performance. After implementing each of the four regression-based methods outlined above on the time series data, we re-implemented them using the ADAM approach.

### 2.5.3 Feature-based methods

The feature-based approach to anomaly detection can make use of multiple time series to identify observations that deviate by distance or density from the majority of data in high dimensional 'feature space' (e.g. Talagala et al., 2018; Wilkinson, 2018). In the initial phase, transformations (e.g. log or differencing transformations) are applied to multiple time series to highlight different anomalies, such as sudden spikes and shifts. Different unsupervised anomaly detection methods are then applied to the high dimensional data space constructed by the transformed series to classify the anomalies. Because feature-based methods take the correlation structure of multiple water-quality variables into account, the anomaly classifications have a probabilistic interpretation. In other words, the anomalous threshold is not a user-defined parameter, but is instead determined by the data using probability theory. This increases the generalisability of such methods across different applications. Feature-based methods are computationally efficient and as such are suitable for analysing big data in near-real time. In addition, they are unsupervised, data-driven approaches and therefore do not require training (Table S1). Here, we implemented HDoutliers (Wilkinson, 2018), aggregated *k*-nearest neighbour (*k*NN-agg; Angiulli and Pizzuti, 2002; Madsen, 2018) and summed *k*-nearest neighbour AD (*k*NN-sum; Madsen et al. 2018) on one set of multivariate data for each site: the turbidity and conductivity time series.

The HDoutliers algorithm proposed by Wilkinson (2018) defines an anomaly as an observation that deviates markedly from the majority by a large distance in high-dimensional space. The algorithm starts by normalizing each time series to prevent variables with large variances having disproportional influence on Euclidean distances. The method uses the Leader algorithm (Hartigan, 1975) to identify anomalous clusters from which a representative member is selected. Nearest neighbour distances between the selected members are then calculated and form the primary source of information for the AD process. An extreme-value theory approach is used to calculate an anomalous threshold, which thus has a probabilistic interpretation.

The HDoultiers algorithm considers only the nearest neighbour distances to identify anomalies. Following Angiulli and Pizzuti (2002), Madsen (2018) proposed an algorithm using *k* nearest neighbour distances. For each observation, the *k*-nearest-neighbours (*k*NN) are first identified using a *k*-dimensional tree (*k*d-tree; Bently, 1975) and an anomaly score is then calculated based on the distances to those neighbours. While *k*NN-agg computes an aggregated distance to the *k*NN (see below), *k*NN-sum simply sums the distances to the *k*NN. The aggregated distance is calculated by aggregating the results from *k*-minimum-nearest neighbours (*k*minNN) to *k*-maximum nearest neighbours (*k*maxNN), such that if *k*minNN = 1 and *k*maxNN = 3, the results from 1NN, 2NN and 3NN are aggregated by taking the weighted sum, assigning nearest neighbours higher weights relative to the neighbours farther apart. Here, we used $k = 10$, the maximum default value of $k$ in Madsen (2018) because it is a suitable trade-off between too low or high a value that may influence performance adversely (McCann and Lowe, 2012).

## 2.6 Select metrics to evaluate and compare methods (Step 6)

We selected several metrics to evaluate and compare the ability of the different AD methods outlined in Section 2.5, to detect the anomalies identified and labelled in Step 3 (Section 2.3), at the different sites for the different water-quality variables, anomaly classes and types (Table 2; Figures S2-S3). We included commonly used metrics calculated easily from the confusion matrix of true and false positives and true and false negatives (TP, FP, TN, FP, respectively; Table S2). These included accuracy and error rate along with two metrics designed to better capture the performance of methods when the number of anomalous versus 'normal' observations is unbalanced, specifically the negative and positive predictive values (NPV and PPV, respectively; Ranawana and Palade, 2006). Finally, we used the root mean square error (RMSE) from the regression-based methods as an additional metric of performance for those methods.

Computation time can also provide insight on the comparative performance of methods. Both regression- and feature-based methods take time for classification. However, feature-based methods classify the complete time series in a single run. By contrast, regression-based methods require additional time for training for prediction, with the exception of naïve methods. Regression-based methods (barring naïve prediction) also require additional time to perform optimization to estimate the model parameters; whilst this can be relatively fast for linear models, non-linear optimization is more time consuming. For these reasons, we can state *a priori* that running the feature-based methods will require less computational time than the regression-based methods. Furthermore, HDoutliers requires less computational time than both *k*NN methods because the former considers only the single most-nearest neighbour whereas the latter consider all *k* nearest neighbours. However, if the feature-based methods were to be implemented in near-real time to classify the time series with newly measured observations, this would make them more computationally comparable with regression-based methods, which are implemented in a loop that forecasts and classifies the one-step-ahead observation as anomalous or otherwise. As such, any difference in classification times between the approaches will depend on the models fitted and the features computed.

## 2.7 Prepare data for anomaly detection (Step 7)

Class 2 anomalies (i.e. impossible values of type F, out-of-sensor-range of type G and missing data of type K) were detected by the automated, hard-coded, classification rules in near-real time. For other anomalies, we implemented regression-based or feature-based methods. To prepare the 'clean' training data for the regression-based AD, we removed all the labelled anomalies from the time series data (Classes 1 and 3). Regression-based AD then followed using the natural log-transformed 'clean' time series for training and the natural log-transformed original times series for testing, for all methods except for linear autoregression for which we took the differences of the natural logarithms. These transformations were applied to meet assumptions of the regression models; forecasting was performed on the transformed scale. Where zero (e.g. type F anomalies in conductivity at PR) or negative values (e.g. type F anomalies in conductivity at PR and in level at SC) were present, we replaced each value with the (non-zero, positive) value of the previous observation to enable forecasting. To calculate the confusion-matrix based performance metrics for the regression-based methods, we first summed the 100% correctly detected Class 2 anomalies to the true positive (TP) count from the regression method before calculating the rest of the metrics (Table S2).

For feature-based AD, we applied both the one-sided and the two-sided derivative transformations to the natural log-transformed turbidity and conductivity time series because exploratory analyses indicated that these transformations highlighted the priority type A anomalies (e.g. large sudden spikes, Class 1) well in feature space. For the one-sided transformation, we took the negative side of the derivative for turbidity, and the positive side for conductivity. Feature-based AD on the transformed time series then followed. We followed the same process as for the regression-based methods, regarding the TP count, to calculate the complete set of confusion matrix-based performance metrics.

## 2.8 Implement anomaly detection methods (Step 8)

We used the *forecast* package (Hyndman, 2017) to implement the regression-based AD methods and the *DDoutliers* package (Madsen, 2018) run within the *oddwater* package (Talagala and Hyndman, 2018) to implement the feature-based AD methods in R statistical software (R Core Team, 2017). We used the same rule-based code to implement the automated classification rules within the regression- and feature-based methods. The R code for the automated classification rules and regression-based methods is provided in the Supplementary materials, along with files containing the time series data and anomaly-type coding. Madsen (2018) and Talagala and Hyndman (2018) describe the R code to implement the feature-based methods described herein.

# 3. Results

## 3.1 Anomalies and their types

Overall, we labelled 1651 turbidity, 521 conductivity and 8 level observations as anomalous in the time series data (Table 2). The majority of these anomalies were of type E (comprising periods of anomalous high variability), H (drift) and L (other).

There was imbalance in the number of non-anomalous (many) to anomalous (few) data points in the time series we used, as well as different types of anomalies (e.g. many type L vs few type A; Table 2). Furthermore, some anomaly types comprised multiple observations (e.g. other type L, drift type H) where as others contained only one (e.g. a type A anomaly). Such imbalances need to be considered in addition to the anomaly-type priority rankings when comparing and interpreting the performance of different methods with respect to their abilities to detect the different anomaly types.

The turbidity time series contained the most anomalies, at both SC and PR, followed by conductivity at PR. There were no clear examples of type C (constant offsets), although data labelled as L (other) between points of sudden shift may have been due to calibration errors manifesting as offsets. In addition, there were no examples of type G anomalies (out-of-sensor-range values). However, there were numerous impossible values (type F), which can be detected by automated classification rules in the same way as type G anomalies. Clusters of spikes (type I) and periods of low variability or persistent values (type B) were also absent. Type K anomalies (missing data) were present in all of the time series.

## 3.2 Evaluate and compare anomaly detection methods (Step 9)

We evaluated and compared results of the various AD methods as part of Step 9 of the AD Framework (Figure 1), as outlined below.

### 3.2.1 Automated classification rules

As expected, the automated classification rules detected all of the Class 2 anomalies (types F, G and K; Table 2) correctly, with no false positives, in each of the time series.

### 3.2.2 Regression-based methods

Results of the regression-based methods performed on the turbidity and conductivity time series at both PR and SC indicated that, in general, all methods had high accuracy (values > 0.80) and low error rates (< 0.20), except when ADAM was used (Table 3). ADAM was associated with high rates of false positive detection (i.e. incorrect classification of normal observations as anomalies), which negatively affected the accuracy and error rates (Figures 3-4 and S4-S9). For example, naïve prediction with ADAM applied to the turbidity time series at PR classified over 5000 observations as false positives compared to 133 without mitigation using AD alone (Table 3, Figure 3). In many cases, large contiguous numbers of false positives occurred when the observations subsequent to a classified anomaly did not display ‘normal’ behaviour relative to the observations classified most recently as non-anomalous. Despite this drawback, ADAM was useful for correct classification of Class 3 anomalies where AD alone was not. For example, 718 out of 718 and 713 out of 915 type E (high variability) anomalies in the turbidity time series at PR and SC, respectively, were detected by naïve ADAM, and all 397 Type H (drift) and 80 type L (other) anomalies in the conductivity time series at PR were detected by ARIMA ADAM (Table 4). ADAM was also useful for detection of anomalous observations that proceeded sudden shifts, such as the L type anomalies in the middle of the turbidity time series (Figures S2 and 3-4).

RegARIMA did not outperform ARIMA, despite the additional water-quality data that were used as covariates. This was especially true for conductivity at PR, where inclusion of other water-quality variables as covariates greatly reduced the rate of correct classification (RegARIMA PPV of 0.49 vs ARIMA PPV of 0.93; Table 3). This likely reflects the characteristics of the water-quality time series at this site, with conductivity displaying complex relationships with both turbidity and level (Figure 2). Thus, including these covariates had a detrimental impact on classification performance. In addition, the behaviour of conductivity tended to be more stable than turbidity through time, somewhat reflective of random walk behaviour, on which naïve prediction (ARIMA(0,1,0)) is based (Hyndman and Koehler, 2006). This may be why the ARIMA(3,1,2) model performed similarly well to naïve prediction when applied to the conductivity time series at PR, given both were using a difference (*d*) parameter of 1 (Table 3, Figures 3-4).

There were only two observations labelled as anomalies in the conductivity time series at SC, and both were of Class 1 (one sudden spike A and one sudden shift D). These two anomalies were classified correctly by all methods, with zero false negatives (Table 4, Table S4). However, all of the methods classified many 'normal' observations incorrectly as anomalies (false positives), particularly ADAM (up to 5091 out of 6280 observations; Table S4), as was the case for other time series at both SC and PR (Table 3). Due to the heavy imbalance of normal versus anomalous observations in the conductivity data at SC, we decided not to undertake further interpretation of the regression-based performance metrics for this time series.

Diagnostics conducted on the residuals of each regression-based method (Figures S13-20) indicated heteroscedasticity was present. In other words, there was change in variance of the data through time (a form of nonstationarity), despite the transformations applied to the time series. Although this will not bias the model forecasts, it may have reduced the accuracy of the prediction intervals and hence affected the classification of anomalies. There was also evidence of nonstationarity in terms of non-constant means in the PR turbidity and conductivity residuals from the linear autorgression, ARIMA and RegARIMA and in the SC turbidity residuals from the ARIMA and RegARIMA models (Box-Ljung tests, $p < 0.05$).

### 3.2.3 Feature-based methods

Each feature-based method applied to the turbidity time series at PR had the same accuracy (0.88), error rate (0.12) and NPV score (0.88; Table 5; Figure 5). The *k*NN-agg method applied to the derivatives of the time series correctly classified the most anomalies (6) of all feature-based methods applied to the PR turbidity data, but also resulted in the most false positives (7) and thus the lowest NPV score (0.46). The HDoutliers method applied to the derivatives of the time series attained the highest NPV score of 0.75, thus attaining the highest values of both NPV and PPV. All methods had high rates of false negative detection (> 720; Table 5), which were associated predominantly with poor detection of Class 3 anomalies; none of the 718 type L ('other') anomalies were classified as such in the turbidity times series by any feature-based method; Table 6, Figure 4). Furthermore, only the methods applied to the derivatives of the turbidity time series correctly classified the type A (sudden spike) and one of the type D (sudden shift) anomalies (Table 11).

For conductivity at PR, accuracy was high (0.92) and error rate was low (0.08; Table 5; Figure S10). The PPV values were all identical and high (0.92), with slightly more variability in the NPV scores, which were also high (0.93 – 0.98); the *k*NN-sum method on the one-sided derivatives attained the highest NPV. However, the feature-based methods tended to produce high false negative rates for the conductivity data, as was the case with the turbidity data at PR. Most methods were able to correctly classify the type A and D anomalies (Table 6).

For turbidity at SC, accuracy (0.83) and error rate (0.17) were the same for each method, as was the case for turbidity at PR (Table 5; Figure S11). NPV scores ranged from 0.42, attained by the *k*NN-agg method, to 0.75 attained by HDoutliers, both of which were applied to the one-sided derivatives of the time series (Table 5). All methods had high false negative rates (> 900; Table 5), but all methods classified three of the four type A anomalies correctly (Table 6).

For the feature-based methods applied to the conductivity time series at SC, we followed the same protocol as we did for the regression-based methods (Table S5, Figure S12), keeping interpretation to a minimum given there were only two anomalies labelled in these data. All methods classified one true positive only (Type A) and misclassified the other anomaly (type D) as normal, but most non-anomalous observations were classified correctly as true negatives (Tables 6 and S5).

# 4. Discussion

The final step of the AD Framework (Figure 1) involves making recommendations based on the results of the different AD methods applied. Here, results of the regression-based methods indicated that the ARIMA method may be useful for AD in streaming water-quality data because it encompasses both naïve prediction (ARIMA(*0,1,0*)) and linear autoregression models (ARIMA(*p,0,0*)) within its suite of possible models. Furthermore, ARIMA may be particularly useful when no other covariates are available to include in RegARIMA models, relationships among potential covariates are complex, such as at PR, or covariates contain missing values. Regarding decisions on whether to include anomaly mitigation as well as detection, ARIMA without mitigation (i.e. without ADAM) may be most useful when the end-user focus is on detection of Class 1 anomalies (sudden spikes and shifts). Such anomalies, if not detected and accounted for, are likely to incorrectly inflate or deflate summary statistics (e.g. monthly means) used in water quality assessments and for compliance checking by water management agencies.

ARIMA with mitigation (i.e. with ADAM) could be implemented subsequently or alternatively to ARIMA alone to detect Class 3 anomalies (e.g. drifts, periods of high variability). Occurrence of such anomalies can indicate that sensors need re-calibrating, and their detection would be of particular value in terms of sensor maintenance. ARIMA models assume that observations are evenly spaced in time, which may become problematic for the models, specifically for the characteristics of the training datasets, if *in situ* water-quality measurements become less frequent and/or irregular in time. This may be especially problematic in training datasets if natural water-quality events are missed. However, increasing the frequency of measurements during high-flow events to capture greater resolution in water-quality dynamics is less problematic. Most importantly, the training dataset should include the full range of natural water-quality dynamics.

Regression-based methods of AD are semi-supervised, and as a result are influenced strongly by the training data used to build the models. In this case study, high rates of false positives were detected in the water-quality time series when these methods of AD were used (Table S1). Yet, decisions on how to dissect time series data into training and test components are not trivial, particularly when there are time-specific behaviours in the data such as seasonality of events and/or when the time series are of limited length (e.g. one year, as was the case here). Methods such as event-based cross validation (Lessels and Bishop, 2013) and walk-forward cross-validation (Bergmeir et al., 2018) may provide potential solutions that could be implemented in future research.

In our analysis, the regression models may have been over-fitted because they were trained on the same data (minus anomalies) used for testing. Using training data from another nearby site on the same watercourse or from a different time period at the same site could lessen this issue. However, there were no nearby sites on PR or SC from which water-quality data from *in situ* sensors were available. If such data become available in the future, training could be performed on those data to see if the AD performance changes.

In rivers, water-quality patterns through time often change with the flow regime (Poff et al., 1997; Nilsson and Renöfält, 2008). This is particularly apparent in highly seasonal rivers such as those of Australia's tropical north, where water quality tends to fluctuate more rapidly and to a greater extent during high-flow events in the wet season than during the more stable low-flow phase of the dry season (Leigh, 2013; O'Brien et al., 2017). This can manifest as nonstationarity in the water-quality time series; for example, as changing variance through time, as was the case here. As such, differentiating between regimes for training purposes may additionally improve the performance of regression-based AD methods in water-quality time series from such rivers. For example, discrete-space hidden Markov models could be used to classify (i.e. segment) the time series into a subset of water-quality regimes found in the data. The regression-based models that require a training dataset (Table S1) could then be applied subsequently to each of the segmented datasets.

Like the regression-based methods without ADAM, the feature-based methods we implemented were not proficient at detecting Class 3 anomalies. This is not surprising given the transformations and algorithms used to implement these methods were developed specifically to prioritize detection of Class 1 anomalies, as per the end-user needs and goals in our case study. Other transformations of the time series data may be required to better target Class 3 anomalies using feature-based methods. This should be borne in mind when transferring our approach to other applications and end-user objectives, such as the monitoring and detection of security intrusions (Teodoro et al. 2009; Talagala et al. 2018). Furthermore, whilst HDoutliers is more computationally efficient than *k*NN methods of feature-based AD, *k*NN methods may be preferable when clusters of anomalies are present in the high-dimensional feature-space produced from the transformed time-series data. Such clusters could manifest if, for example, there were several sudden spikes in the time series, each of the same value. Such phenomena may result from recurrent technical issues with the sensor equipment that produce a specific, recurrent anomalous value.

# 5. Conclusions

Our results highlight that a combination of methods, as recommended in Section 4, is likely to provide optimal performance in terms of correct classification of anomalies in streaming water-quality data from *in situ* sensors, whilst minimizing false detection rates. Furthermore, our framework emphasizes the importance of communication between end-users and anomaly detection developers for optimal outcomes with respect to both detection performance and end-user application. To this end, our framework has high transferability to other types of high frequency time-series data and anomaly detection applications. Within the purview of water-quality monitoring, for example, our framework could be applied to other water-quality variables measured by *in situ* sensors that are used commonly in ecosystem health assessments, such as dissolved oxygen, water temperature and nitrate (Leigh et al., 2013; Pellerin et al., 2016). These properties of water are highly dynamic in space and time (Hunter and Walton, 2008; Boulton et al., 2014) and so differentiating anomalies from real water-quality events may be more challenging than it is for properties like turbidity and conductivity investigated in this study. These latter two properties hold promise as near-real time surrogates of sediment and nutrient concentrations (Jones et al., 2011; Slaets et al., 2014), which would reduce the amount of laboratory analysis otherwise required for discrete water samples. Therefore, the extension of automated AD methods, as developed herein, along with models that predict sediment and nutrient concentrations from these data, into space and time on river networks, could revolutionize the way we monitor and manage water quality, whilst also increasing scientific understanding of the spatio-temporal dynamics of water-quality in rivers and the potential effects on downstream waters.

# Acknowledgements

Funding for this project was provided by the Queensland Department of Environment and Science (DES) and the ARC Centre of Excellence for Mathematical and Statistical Frontiers (ACEMS). A repository of the water-quality data from the *in situ* sensors used herein and the code used to implement the regression-based anomaly detection methods are provided in the Supplementary materials.

**Table 1:** Types of anomalies likely encountered in *in situ* sensor-generated water-quality time series, along with the importance ranking of each type with respect to the priority end-user goal in this case study (i.e. time series visualization), and relevance to potential end-users.

| Anomaly type | Type code (Class†) | Description | Examples in the literature and/or alternative terminology | Importance ranking (with respect to time series visualization in this case study) | Potential end-users and applications‡ |
|---|---|---|---|---|---|
| Large sudden spike | A* (1) | Anomalous value is isolated and 'much' higher or lower than surrounding data, and the spike occurs in a very short window of time (e.g. only one data point is anomalously high or low). | Point or collective anomaly (Goldstein and Uchida 2016) | First priority (at any point in the time series) | Management, monitoring and compliance; Policy and decision makers; Public; Data managers; Sensor maintenance technicians |
| Low variability / persistent values | B (3) | Values constant though time or with very minimal variation compared with that expected | Data value persistence (Horsburgh et al., 2015); collective anomaly (Chandola et al., 2009) | Second priority (especially during event flow) | Data managers; Sensor maintenance technicians |
| Constant offset (e.g. calibration error) | C (3) | Values are in error by some constant. Likely related to / seen before and/or after sudden shifts | Incorrect offset or calibration (Horsburgh et al., 2015) | Third priority | Data managers; Sensor maintenance technicians |
| Sudden shifts | D (1) | Values suddenly shift to a new range (higher or lower than previous range) | Level shifts (Tsay, 1988) | Equal third priority (especially when shift is considered large) | Management, monitoring and compliance; Policy and decision makers; Public; Data managers; Sensor maintenance technicians |
| High variability | E (3) | Values oscillate considerably over short time periods (more than expected during natural daily cycles or events) | Variance change (Tsay, 1988); collective anomaly (Chandola et al., 2009) | Fourth priority | Data managers; Sensor maintenance technicians |
| Impossible values | F (2) | Values impossible or highly unlikely for that water-quality variable (e.g. negative values for all, conductivity values nearing or at zero ('too fresh')) | Out of range values (Horsburgh et al., 2015) | Important, but should be detected easily (e.g. using a simple rule) | Sensor manufacturers; Statisticians; Data managers; Sensor maintenance technicians |
| Out-of-sensor-range values | G (2) | Values that the sensors are incapable of detecting (outside of their detection range). Some of these anomalies may be first captured under type F (impossible values) | Not distinguished from type F by Horsburgh et al. (2015) | Important, but should be detected easily (e.g. using a simple rule) | Sensor manufacturers; Statisticians; Data managers; Sensor maintenance technicians |
| Drift | H (3) | Gradual change in values in positive or negative direction | Sensor drift (Horsburgh et al., 2015); collective anomaly (Chandola et al., 2009) | Comparatively low priority (most likely observed in turbidity), but important to flag as being a possible occurrence of an anomaly e.g. when gradual increase or decrease occurs before a sudden shift | Sensor manufacturers; Data managers; Sensor maintenance technicians |
| Clusters of spikes | I* (1) | Multiple spikes in a short period of time | Micro cluster (Goldstein and Uchida 2016); collective anomaly (Chandola et al., 2009) | Low priority (isolated spikes much more important to detect) | Management, monitoring and compliance; Policy and decision makers; Public; Data managers; Sensor maintenance technicians |
| Small sudden spike | J* (1) | Anomalous value is 'somewhat' higher or lower than surrounding data, and the spike occurs in a very short window of time (e.g. only one data point is anomalously high or low) | Point anomaly (Goldstein and Uchida 2016) | Very low priority | Data managers; Sensor maintenance technicians |
| Missing values | K (2) | Gaps in time series (i.e. greater than the set frequency of measurement) | Skipped or no-data values (Horsburgh et al., 2015) | Undetermined | Data managers; Sensor maintenance technicians; Sensor manufacturers; Statisticians; Policy and decision makers |

*Spikes may be in the positive or negative direction with respect to surrounding data (i.e. can include a sudden isolated decrease and/or a sudden isolated increase in value). †Classes of anomalies, as defined in this paper: (1) involve a sudden change in value from the previous observation, (2) are detectable by automated classification rules, (3) likely require user intervention to identify observations as anomalous. ‡Monitoring, management and compliance: agencies, industries and landholders etc. concerned with water quality monitoring, management and compliance checking – summary statistics such as means are strongly influenced by such anomalies; Policy and decision makers – to limit use of incorrect data and for reporting purposes; Public – to avoid false warning

of water quality breaches; Data managers – for quality control and assurance and to increase confidence in the data by reporting the presence of such anomalies; Sensor maintenance technicians – to ensure timely and correct calibration and maintenance of equipment; Sensor manufacturers – to improve performance, e.g. extend battery life, improve wiper quality to further minimise biofouling; Statisticians – for AD methods to better detect other non-trivial anomaly types and/or for methods requiring regular and frequent observations.

**Table 2:** Number of anomalous observations identified according to type, class and water-quality variable at Pioneer River (PR) and Sandy Creek (SC). Number of instances of Class 3 anomalies that comprise multiple contiguous observations, and where relevant their neighbouring anomaly types, in parentheses.

| Site | Anomaly type and class | Turbidity | Conductivity | Level | Total |
|---|---|---|---|---|---|
| PR | A (Class 1) | 1 | 2 | 0 | 3 |
| | D (Class 1) | 3 | 2 | 0 | 5 |
| | F (Class 2) | 0 | 34 | 0 | 34 |
| | H (Class 3) | 0 | 397 (1 instance, before a D) | 0 | 397 |
| | J (Class 1) | 5 | 0 | 2 | 7 |
| | K (Class 2) | 4 | 4 | 4 | 12 |
| | L (Class 3) | 718 (1 instance, between two Ds) | 80 (2 instances, the first after a D, the second between Ks) | 0 | 798 |
| | Class 1 | 9 | 4 | 2 | 15 |
| | Class 2 | 4 | 38 | 4 | 46 |
| | Class 3 | 718 | 477 | 0 | 1195 |
| | *Total (out of 6280 observations)* | *731* | *519* | *6* | *1256* |
| SC | A (Class 1) | 4 | 1 | 0 | 5 |
| | D (Class 1) | 1 | 0 | 0 | 1 |
| | E (Class 3) | 914 (2 instances, the second before a D) | 0 | 0 | 914 |
| | F (Class 2) | 0 | 0 | 1 | 1 |
| | K (Class 2) | 1 | 1 | 1 | 3 |
| | Class 1 | 5 | 1 | 0 | 6 |
| | Class 2 | 1 | 1 | 2 | 4 |
| | Class 3 | 914 | 0 | 0 | 914 |
| | *Total (out of 5402 observations)* | *920* | *2* | *2* | *924* |

**Table 3:** Performance metrics for regression-based methods of anomaly detection performed separately on turbidity and conductivity data from *in situ* sensors at Pioneer River (PR) and Sandy Creek (SC), incorporating 100% detection of Class 2 anomalies by automated classification rules. See Tables S2-3 for metric formulae and descriptions and Section 2.5.2 for model specifics. AD, anomaly detection; ADAM, anomaly detection and mitigation; AR, autoregression.

| **Site** | **Time series** | **Model (*p*,*d*,*q*)** | **Method** | **TN** | **FN** | **FP** | **TP** | **Accuracy** | **Error rate** | **NPV** | **PPV** | **RMSE** |
|---|---|---|---|---|---|---|---|---|---|---|---|---|
| PR | Turbidity | Naïve (0,1,0) | AD | 5416 | 715 | 133 | 16 | 0.86 | 0.14 | 0.88 | 0.11 | 0.21 |
| | | | ADAM | 347 | 0 | 5202 | 731 | 0.17 | 0.83 | 1.00 | 0.12 | 0.21 |
| | | Linear AR (4,0,0) | AD | 5398 | 712 | 151 | 19 | 0.86 | 0.14 | 0.88 | 0.04 | 0.20 |
| | | | ADAM | 4491 | 25 | 1058 | 706 | 0.83 | 0.17 | 0.99 | 0.40 | 0.87 |
| | | ARIMA (3,1,2) | AD | 5405 | 711 | 144 | 20 | 0.86 | 0.14 | 0.88 | 0.12 | 0.20 |
| | | | ADAM | 4465 | 25 | 1084 | 706 | 0.82 | 0.18 | 0.99 | 0.39 | 0.90 |
| | | RegARIMA (5,1,5) | AD | 5344 | 695 | 205 | 36 | 0.86 | 0.14 | 0.88 | 0.15 | 0.57 |
| | | | ADAM | 171 | 0 | 5378 | 731 | 0.14 | 0.86 | 1.00 | 0.12 | 0.39 |
| PR | Conductivity | Naïve (0,1,0) | AD | 5759 | 459 | 2 | 60 | 0.93 | 0.07 | 0.93 | 0.97 | 0.17 |
| | | | ADAM | 4455 | 399 | 1306 | 120 | 0.73 | 0.27 | 0.92 | 0.08 | 0.17 |
| | | Linear AR (2,0,0) | AD | 5709 | 453 | 52 | 66 | 0.92 | 0.08 | 0.93 | 0.56 | 0.17 |
| | | | ADAM | 4256 | 397 | 1505 | 122 | 0.70 | 0.30 | 0.91 | 0.07 | 0.64 |
| | | ARIMA(3,1,2) | AD | 5756 | 455 | 5 | 64 | 0.93 | 0.07 | 0.93 | 0.93 | 0.16 |
| | | | ADAM | 1873 | 0 | 3888 | 519 | 0.38 | 0.62 | 1.00 | 0.12 | 1.37 |
| | | RegARIMA (1,1,2) | AD | 5675 | 437 | 86 | 82 | 0.92 | 0.08 | 0.93 | 0.49 | 0.26 |
| | | | ADAM | 128 | 0 | 5633 | 519 | 0.10 | 0.90 | 1.00 | 0.08 | 0.07 |
| SC | Turbidity | Naïve (0,1,0) | AD | 4386 | 859 | 96 | 61 | 0.82 | 0.18 | 0.84 | 0.39 | 0.24 |
| | | | ADAM | 491 | 134 | 3991 | 786 | 0.24 | 0.76 | 0.79 | 0.16 | 0.24 |
| | | Linear AR (5,0,0) | AD | 4347 | 830 | 135 | 90 | 0.82 | 0.18 | 0.84 | 0.40 | 0.22 |
| | | | ADAM | 2178 | 753 | 2340 | 167 | 0.43 | 0.57 | 0.74 | 0.07 | 1.06 |
| | | ARIMA (3,1,2) | AD | 4348 | 829 | 134 | 91 | 0.82 | 0.18 | 0.84 | 0.40 | 0.22 |
| | | | ADAM | 2187 | 751 | 2295 | 169 | 0.44 | 0.56 | 0.74 | 0.07 | 1.06 |
| | | RegARIMA (5,1,0) | AD | 4345 | 820 | 137 | 100 | 0.82 | 0.18 | 0.84 | 0.42 | 0.23 |
| | | | ADAM | 775 | 81 | 3707 | 839 | 0.30 | 0.70 | 0.91 | 0.18 | 0.06 |

**Table 4:** Number of turbidity (T) and conductivity (C) anomalies of each type and class classified correctly by each regression-based method for Pioneer River (PR) and Sandy Creek (SC). Number of true anomalies and number of instances where relevant indicated in parentheses. Class 2 anomalies detected by automated classification rules. AR, autoregression, - not applicable.

| River (variable) | Model | Method | A<br>Class 1 | D<br>Class 1 | E<br>Class 3 | F<br>Class 2 | J<br>Class 1 | K<br>Class 2 | H<br>Class 3 | L<br>Class 3 |
|---|---|---|---|---|---|---|---|---|---|---|
| PR (T) | | | (1) | (3) | (0) | (0) | (5) | (4) | (0) | (718; 1 instance) |
| | Naïve | AD | 1 | 3 | - | - | 5 | 4 | - | 3 |
| | | ADAM | 1 | 3 | - | - | 5 | 4 | - | 718 |
| | Linear AR | AD | 1 | 3 | - | - | 5 | 4 | - | 6 |
| | | ADAM | 1 | 2 | - | - | 5 | 4 | - | 694 |
| | ARIMA | AD | 1 | 3 | - | - | 5 | 4 | - | 7 |
| | | ADAM | 1 | 3 | - | - | 5 | 4 | - | 694 |
| | RegARIMA | AD | 1 | 3 | - | - | 5 | 4 | - | 23 |
| | | ADAM | 1 | 3 | - | - | 5 | 4 | - | 718 |
| PR (C) | | | (2) | (2) | (0) | (34) | (0) | (4) | (397; 1 instance) | (80; 2 instances) |
| | Naïve | AD | 2 | 1 | - | 34 | - | 4 | 0 | 19 |
| | | ADAM | 1 | 1 | - | 34 | - | 4 | 0 | 80 |
| | Linear AR | AD | 2 | 2 | - | 34 | - | 4 | 0 | 24 |
| | | ADAM | 2 | 2 | - | 34 | - | 4 | 0 | 80 |
| | ARIMA | AD | 2 | 1 | - | 34 | - | 4 | 0 | 23 |
| | | ADAM | 2 | 2 | - | 34 | - | 4 | 397 | 80 |
| | RegARIMA | AD | 2 | 2 | - | 34 | - | 4 | 0 | 40 |
| | | ADAM | 2 | 2 | - | 34 | - | 4 | 397 | 80 |
| SC (T) | | | (4) | (1) | (915; 2 instances) | (0) | (0) | (0) | (0) | (0) |
| | Naïve | AD | 4 | 1 | 276 | - | - | - | - | - |
| | | ADAM | 4 | 1 | 780 | - | - | - | - | - |
| | Linear AR | AD | 4 | 0 | 85 | - | - | - | - | - |
| | | ADAM | 4 | 1 | 161 | - | - | - | - | - |
| | ARIMA | AD | 4 | 0 | 85 | - | - | - | - | - |
| | | ADAM | 4 | 1 | 162 | - | - | - | - | - |
| | RegARIMA | AD | 4 | 1 | 94 | - | - | - | - | - |
| | | ADAM | 4 | 1 | 833 | - | - | - | - | - |
| SC (C) | | | (1) | (1) | (0) | (0) | (0) | (0) | (0) | (0) |
| | Naïve | AD | 1 | 1 | - | - | - | - | - | - |
| | | ADAM | 1 | 1 | - | - | - | - | - | - |
| | Linear AR | AD | 1 | 1 | - | - | - | - | - | - |
| | | ADAM | 1 | 1 | - | - | - | - | - | - |
| | ARIMA | AD | 1 | 1 | - | - | - | - | - | - |
| | | ADAM | 1 | 1 | - | - | - | - | - | - |
| | RegARIMA | AD | 1 | 1 | - | - | - | - | - | - |
| | | ADAM | 1 | 1 | - | - | - | - | - | - |

**Table 5:** Performance metrics for feature-based methods of anomaly detection performed on multivariate water-quality time series from *in situ* sensors at Pioneer River (PR) and Sandy Creek (SC), incorporating 100% detection of Class 2 anomalies by automated classification rules. See Tables S2-3 for metric formulae and descriptions. OS, one sided.

| Site | Time series | Method | Transformation | TN | FN | FP | TP | Accuracy | Error rate | NPV | PPV |
|---|---|---|---|---|---|---|---|---|---|---|---|
| PR | Turbidity | HDoutliers | Derivative | 5548 | 728 | 1 | 3 | 0.88 | 0.12 | 0.75 | 0.88 |
| | Turbidity | | OS Derivative | 5547 | 727 | 2 | 4 | 0.88 | 0.12 | 0.67 | 0.88 |
| | Turbidity | *k*NN-agg | Derivative | 5542 | 725 | 7 | 6 | 0.88 | 0.12 | 0.46 | 0.88 |
| | Turbidity | | OS Derivative | 5546 | 728 | 3 | 3 | 0.88 | 0.12 | 0.50 | 0.88 |
| | Turbidity | *k*NN-sum | Derivative | 5547 | 728 | 2 | 3 | 0.88 | 0.12 | 0.60 | 0.88 |
| | Turbidity | | OS Derivative | 5546 | 728 | 3 | 3 | 0.88 | 0.12 | 0.50 | 0.88 |
| | Conductivity | HDoutliers | Derivative | 5758 | 470 | 3 | 49 | 0.92 | 0.08 | 0.94 | 0.92 |
| | Conductivity | | OS Derivative | 5758 | 479 | 3 | 40 | 0.92 | 0.08 | 0.93 | 0.92 |
| | Conductivity | *k*NN-agg | Derivative | 5759 | 472 | 2 | 47 | 0.92 | 0.08 | 0.96 | 0.92 |
| | Conductivity | | OS Derivative | 5758 | 479 | 3 | 40 | 0.92 | 0.08 | 0.93 | 0.92 |
| | Conductivity | *k*NN-sum | Derivative | 5760 | 471 | 1 | 48 | 0.92 | 0.08 | 0.98 | 0.92 |
| | Conductivity | | OS Derivative | 5759 | 479 | 2 | 40 | 0.92 | 0.08 | 0.95 | 0.92 |
| SC | Turbidity | HDoutliers | Derivative | 4477 | 914 | 5 | 6 | 0.83 | 0.17 | 0.55 | 0.83 |
| | Turbidity | | OS Derivative | 4481 | 917 | 1 | 3 | 0.83 | 0.17 | 0.75 | 0.83 |
| | Turbidity | *k*NN-agg | Derivative | 4477 | 914 | 5 | 6 | 0.83 | 0.17 | 0.55 | 0.83 |
| | Turbidity | | OS Derivative | 4471 | 912 | 11 | 8 | 0.83 | 0.17 | 0.42 | 0.83 |
| | Turbidity | *k*NN-sum | Derivative | 4482 | 920 | 0 | 0 | 0.83 | 0.17 | n/a | 0.83 |
| | Turbidity | | OS Derivative | 4480 | 917 | 2 | 3 | 0.83 | 0.17 | 0.60 | 0.83 |

**Table 6:** Number of turbidity (T) and conductivity (C) anomalies of each type and class classified correctly by each feature-based method for Pioneer River (PR) and Sandy Creek (SC). Number of Pioneer River (PR) turbidity anomalies of each type and class classified correctly by each feature-based method. Number of true anomalies and number of instances where relevant indicated in parentheses. Class 2 anomalies detected by automated classification rules. -, not applicable.

| **River (variable)** | **Method** | **Transformation** | **A Class 1** | **D Class 1** | **E Class 3** | **F Class 2** | **J Class 1** | **K Class 2** | **H Class 3** | **L Class 3** |
|---|---|---|---|---|---|---|---|---|---|---|
| PR (T) | | | (1) | (3) | (0) | (0) | (5) | (4) | (0) | (718; 1 instance) |
| | HDoutliers | Derivative | 1 | 1 | - | - | 1 | 4 | - | 0 |
| | | OS Derivative | 0 | 0 | - | - | 4 | 4 | - | 0 |
| | *k*NN-agg | Derivative | 1 | 1 | - | - | 4 | 4 | - | 0 |
| | | OS Derivative | 0 | 0 | - | - | 3 | 4 | - | 0 |
| | *k*NN-sum | Derivative | 1 | 1 | - | - | 1 | 4 | - | 0 |
| | | OS Derivative | 0 | 0 | - | - | 3 | 4 | - | 0 |
| PR (C) | | | (2) | (2) | (0) | (34) | (0) | (4) | (397; 1 instance) | (80; 2 instances) |
| | HDoutliers | Derivative | 1 | 1 | - | 34 | - | 4 | 0 | 12 |
| | | OS Derivative | 1 | 0 | - | 34 | - | 4 | 0 | 4 |
| | *k*NN-agg | Derivative | 1 | 1 | - | 34 | - | 4 | 0 | 10 |
| | | OS Derivative | 1 | 1 | - | 34 | - | 4 | 0 | 4 |
| | *k*NN-sum | Derivative | 1 | 1 | - | 34 | - | 4 | 0 | 11 |
| | | OS Derivative | 0 | 0 | - | 34 | - | 4 | 0 | 0 |
| SC (T) | | | (4) | (1) | (915; 2 instances) | (0) | (0) | (0) | (0) | (0) |
| | HDoutliers | Derivative | 3 | 0 | 3 | - | - | - | - | - |
| | | OS Derivative | 3 | 0 | 0 | - | - | - | - | - |
| | *k*NN-agg | Derivative | 3 | 0 | 3 | - | - | - | - | - |
| | | OS Derivative | 3 | 0 | 5 | - | - | - | - | - |
| | *k*NN-sum | Derivative | 0 | 0 | 0 | - | - | - | - | - |
| | | OS Derivative | 3 | 0 | 0 | - | - | - | - | - |
| SC (C) | | | (1) | (1) | (0) | (0) | (0) | (0) | (0) | (0) |
| | HDoutliers | Derivative | 1 | 0 | - | - | - | - | - | - |
| | | OS Derivative | 1 | 0 | - | - | - | - | - | - |
| | *k*NN-agg | Derivative | 1 | 0 | - | - | - | - | - | - |
| | | OS Derivative | 1 | 0 | - | - | - | - | - | - |
| | *k*NN-sum | Derivative | 1 | 0 | - | - | - | - | - | - |
| | | OS Derivative | 1 | 0 | - | - | - | - | - | - |

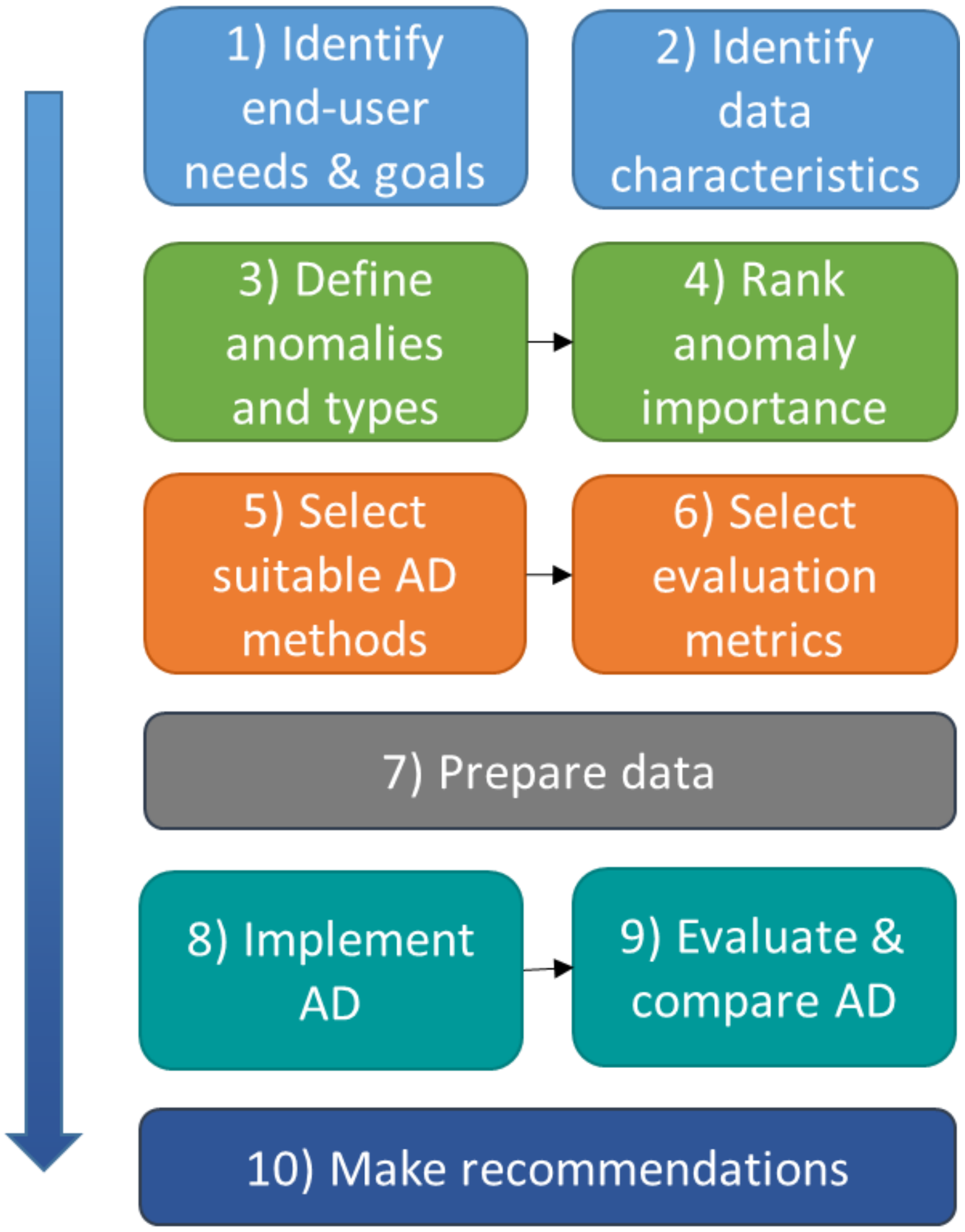

**Figure 1:** The ten-step Anomaly Detection (AD) framework for high frequency water-quality data, which includes defining and ranking the importance of different types of anomalies, based on end-user needs and data characteristics, to inform algorithm choice, implementation, performance evaluation and resultant recommendations. Numbers indicate the order of steps taken. Arrows indicate directions of influence between steps.

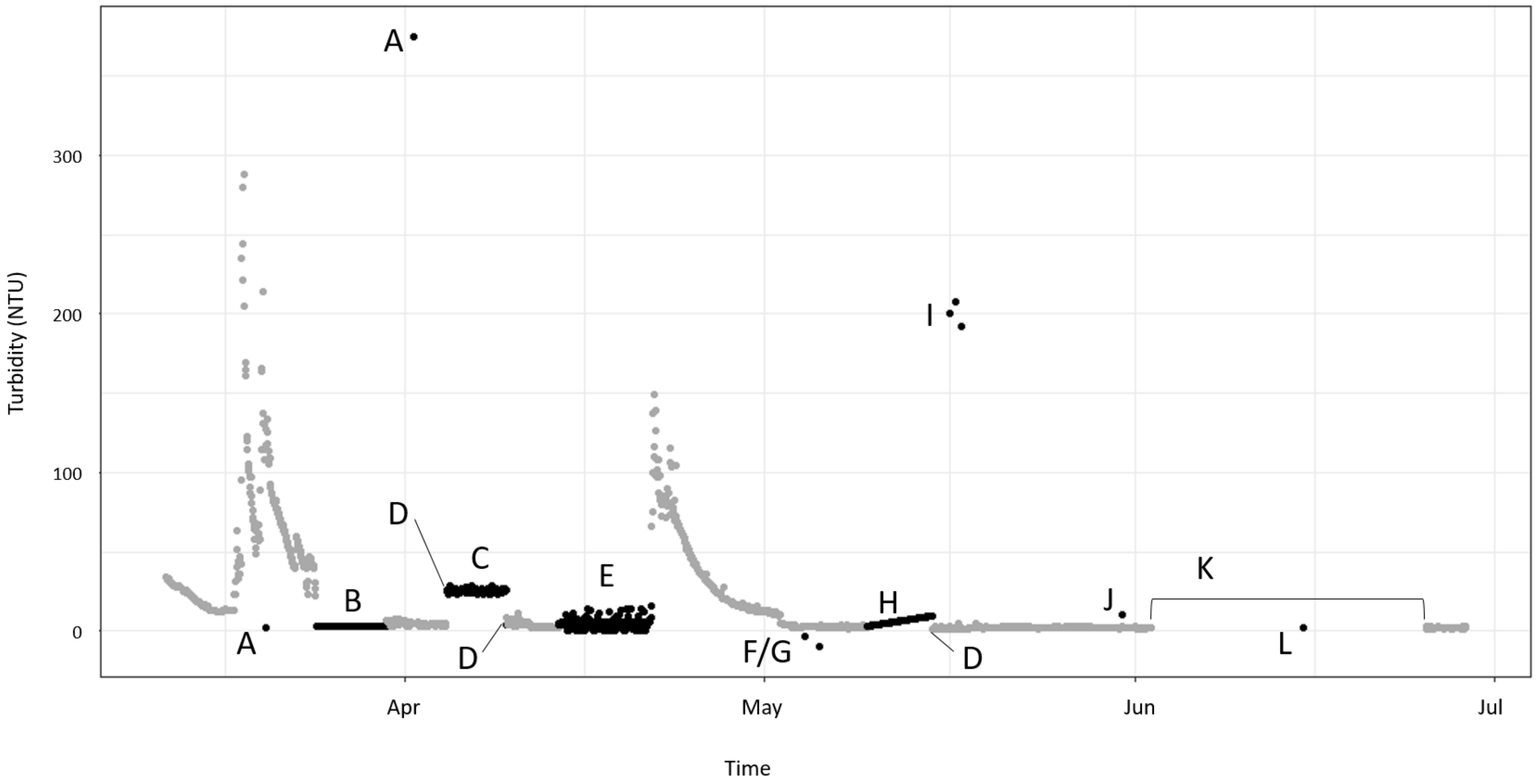

**Figure 2:** Example of a turbidity (NTU) time series featuring both normal observations (dark grey points) and anomalies (black points; labelled A-L according to Table 1).

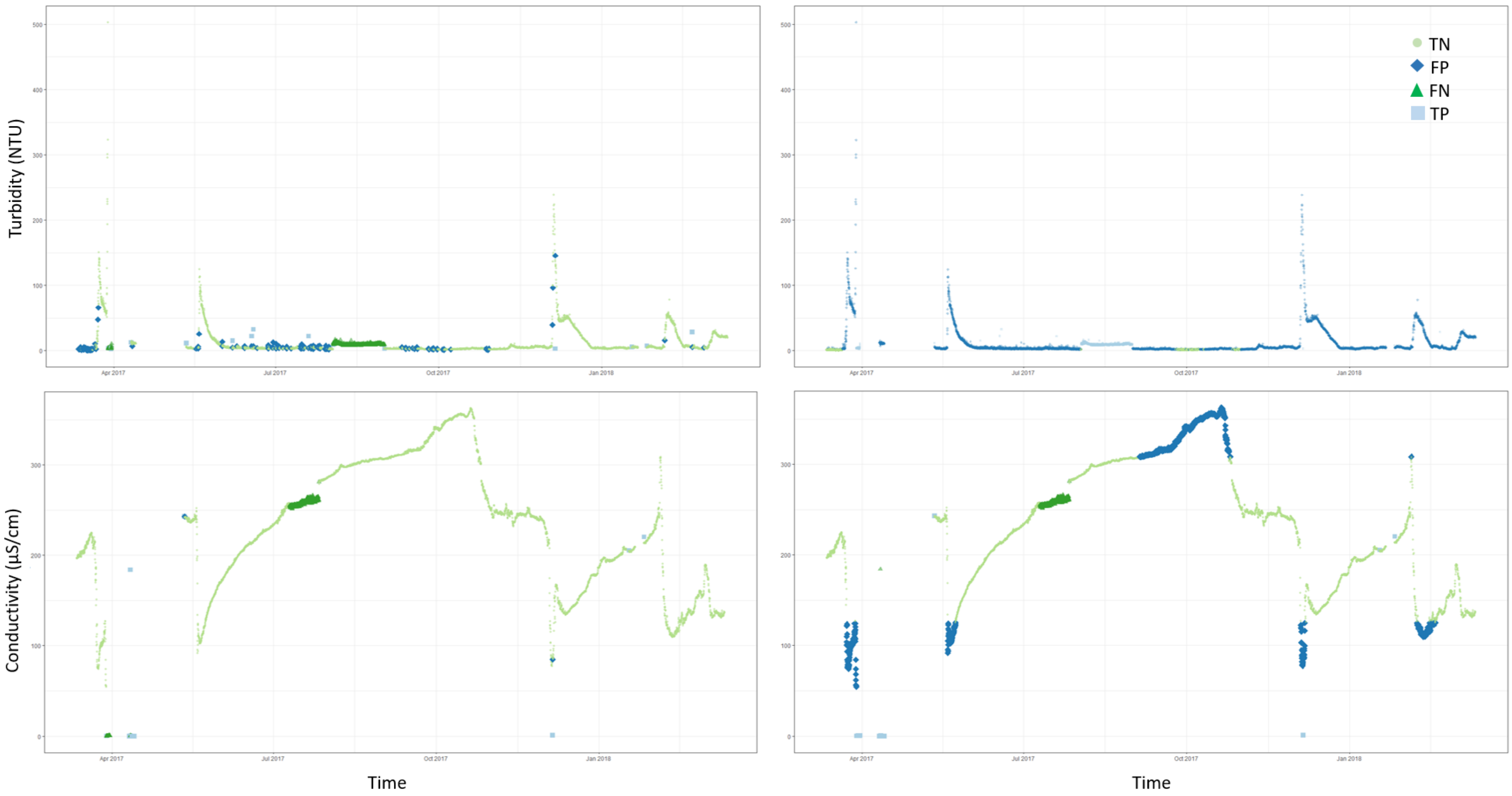

**Figure 3:** Classification of turbidity (upper row) and conductivity observations (lower row) measured by *in situ* sensors at Pioneer River (PR) by naïve prediction as true negatives (TN), false negatives (FN), false positives (FP) or true positives (TP). Plots on the left show results from naïve prediction alone, those on the right show results from naïve prediction with ADAM.

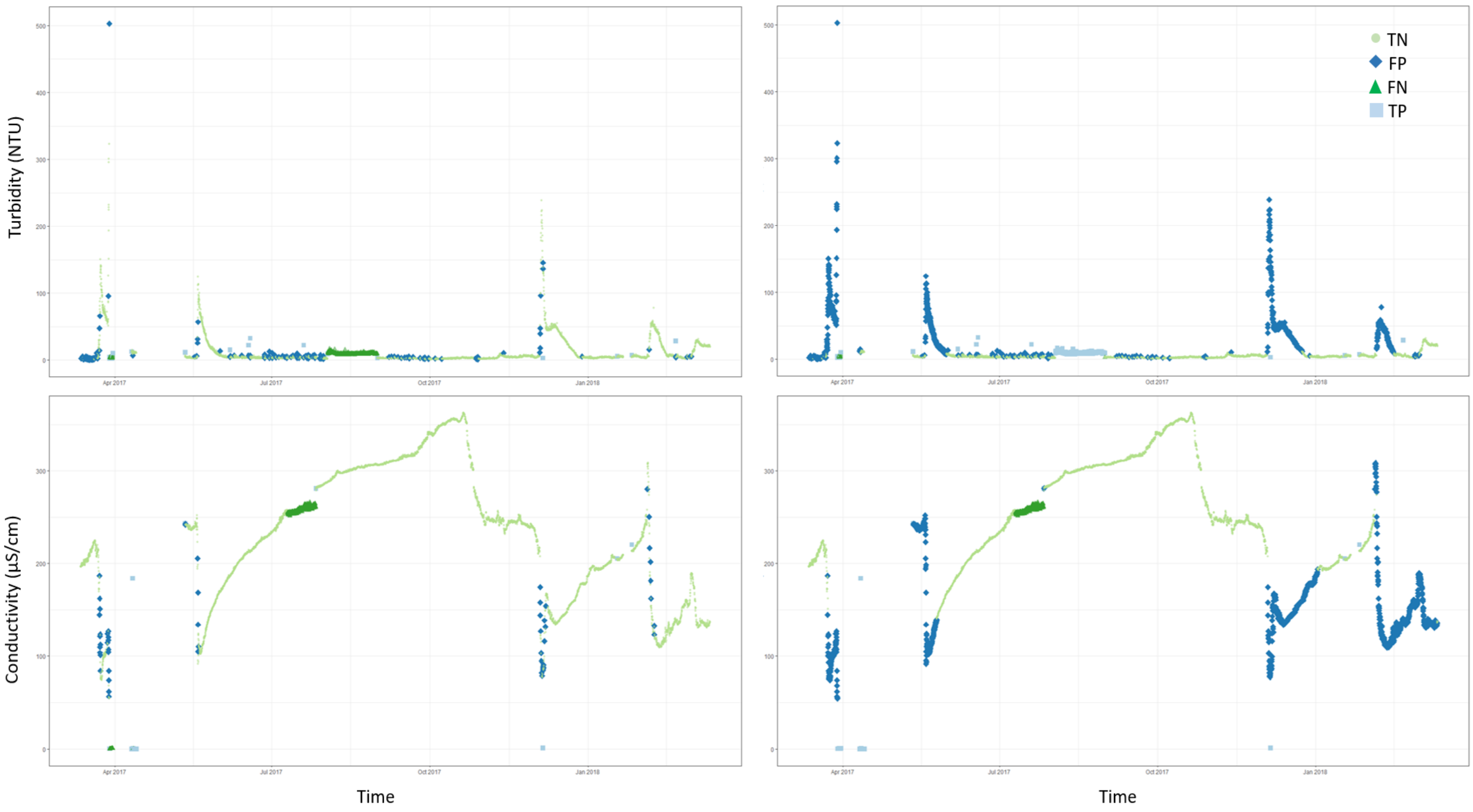

**Figure 4:** Classification of turbidity (upper row) and conductivity observations (lower row) measured by *in situ* sensors at Pioneer River (PR) by ARIMA as true negatives (TN), false negatives (FN), false positives (FP) or true positives (TP). Plots on the left show results from ARIMA alone, those on the right show results from ARIMA with ADAM.

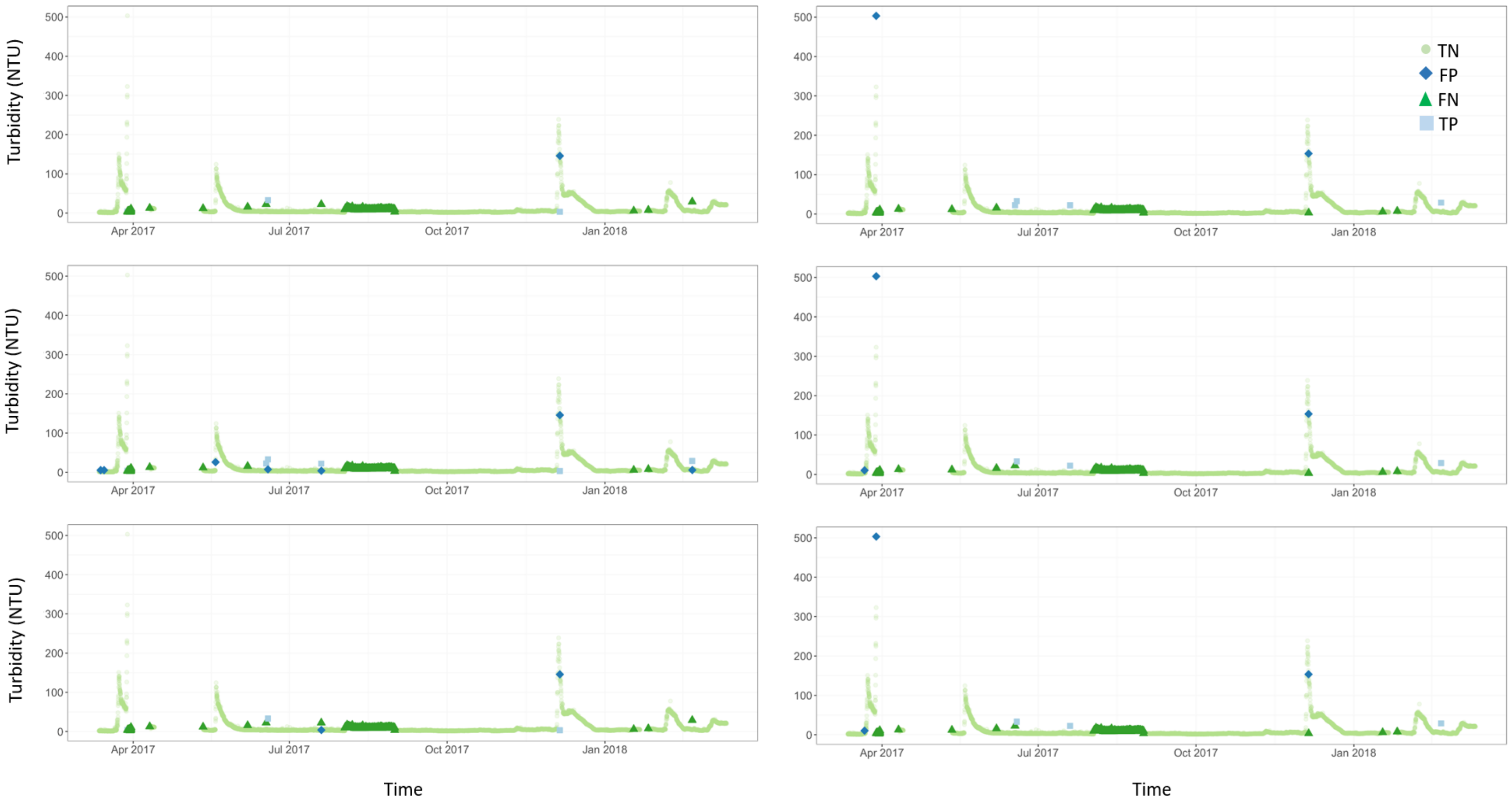

**Figure 5:** Classification of turbidity measured by an *in situ* sensor at Pioneer River (PR) by HDoutliers (upper row), *k*NN-agg (middle row) and *k*NN-sum (lower row) as true negatives (TN), false negatives (FN), false positives (FP) or true positives (TP). Plots on the left show results of methods applied to the derivatives, and those on the right show results of methods applied to the one-sided derivatives of the time series.

# Supplementary materials

The supplementary materials for the article by Leigh et al. "A framework for automated anomaly detection in high frequency water-quality data from in situ sensors" comprise the following:

1. Bivariate relationships and time series showing anomalous and non-anomalous observations for water quality data collected from Pioneer River and Sandy Creek (Figures S1-3)
2. Results of the regression-based and feature-based methods applied to the conductivity time series at Sandy Creek (Tables S1-5);
3. Time series plots not included in the main article that show the observations classified as true positives, false positives, true negatives or false negatives according to each method applied to each time series of water-quality data at each site (Figures S4-S12);
4. Diagnostic plots for the regression-based methods requiring training that were implemented in the main article (Figures S13-20);
5. Files (supplied separately) containing the time series data, supplied by the Queensland Department of Environment and Science (please refer to the Department's website for the disclaimer to these data: https://www.des.qld.gov.au/legal/disclaimer/), and the anomaly-type coding used in the main article (data_pioneer.csv and data_sandy.csv); and
6. Files (supplied separately) containing the R code used to implement the regression-based methods on the time series data, and to calculate performance metrics, as outlined in the main article (PioneerRiver.R, SandyCreek.R, NaivePredictor.R, Prediction.R and ADPerformance.R).

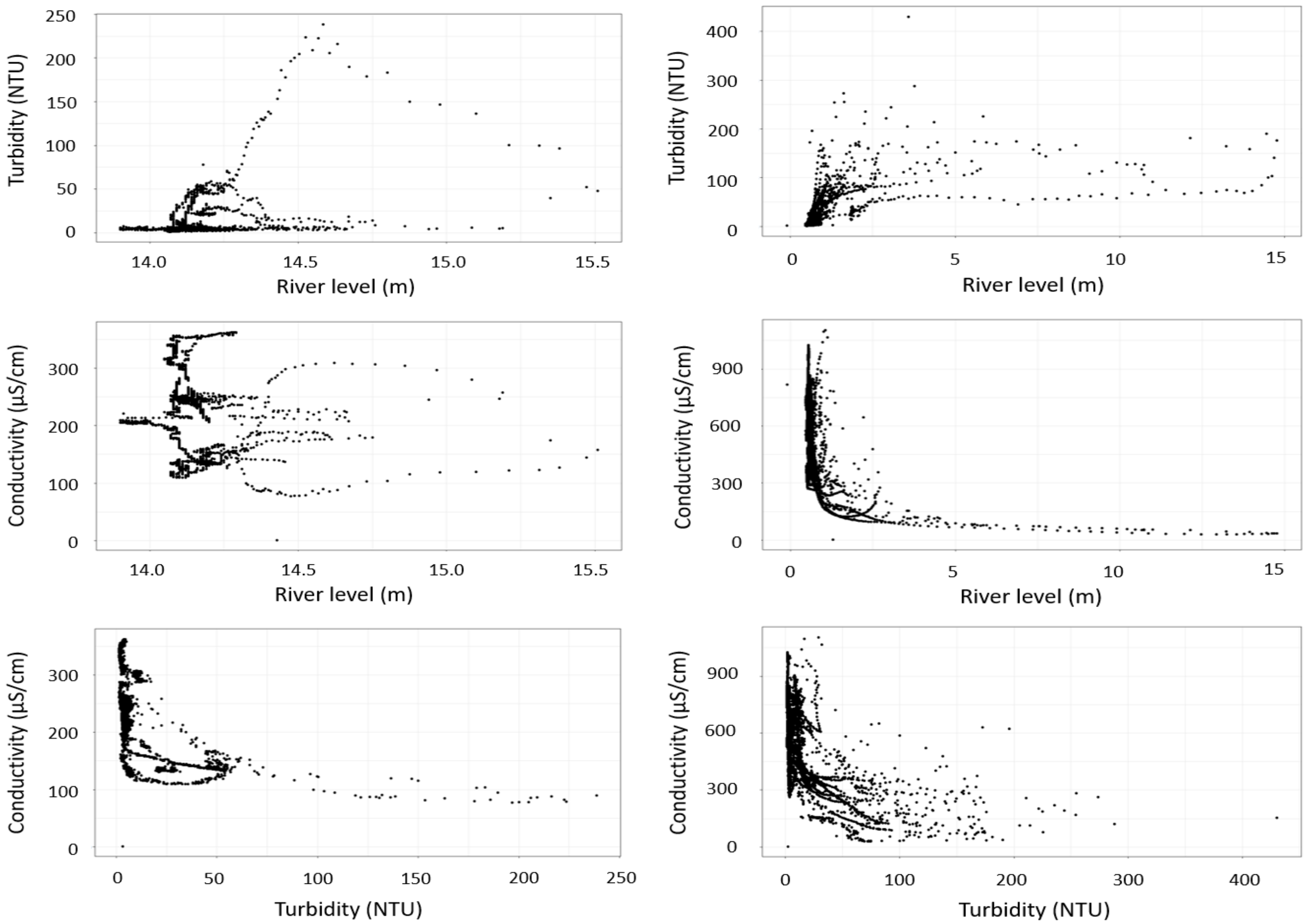

**Figure S1:** Bivariate relationships between the water-quality data (turbidity, NTU; conductivity, µS/cm; river level, m) measured by *in situ* sensors at Pioneer River (PR, left column) and Sandy Creek (SC, right column), including all observations recorded between 12 March 2017 and 12 March 2018.

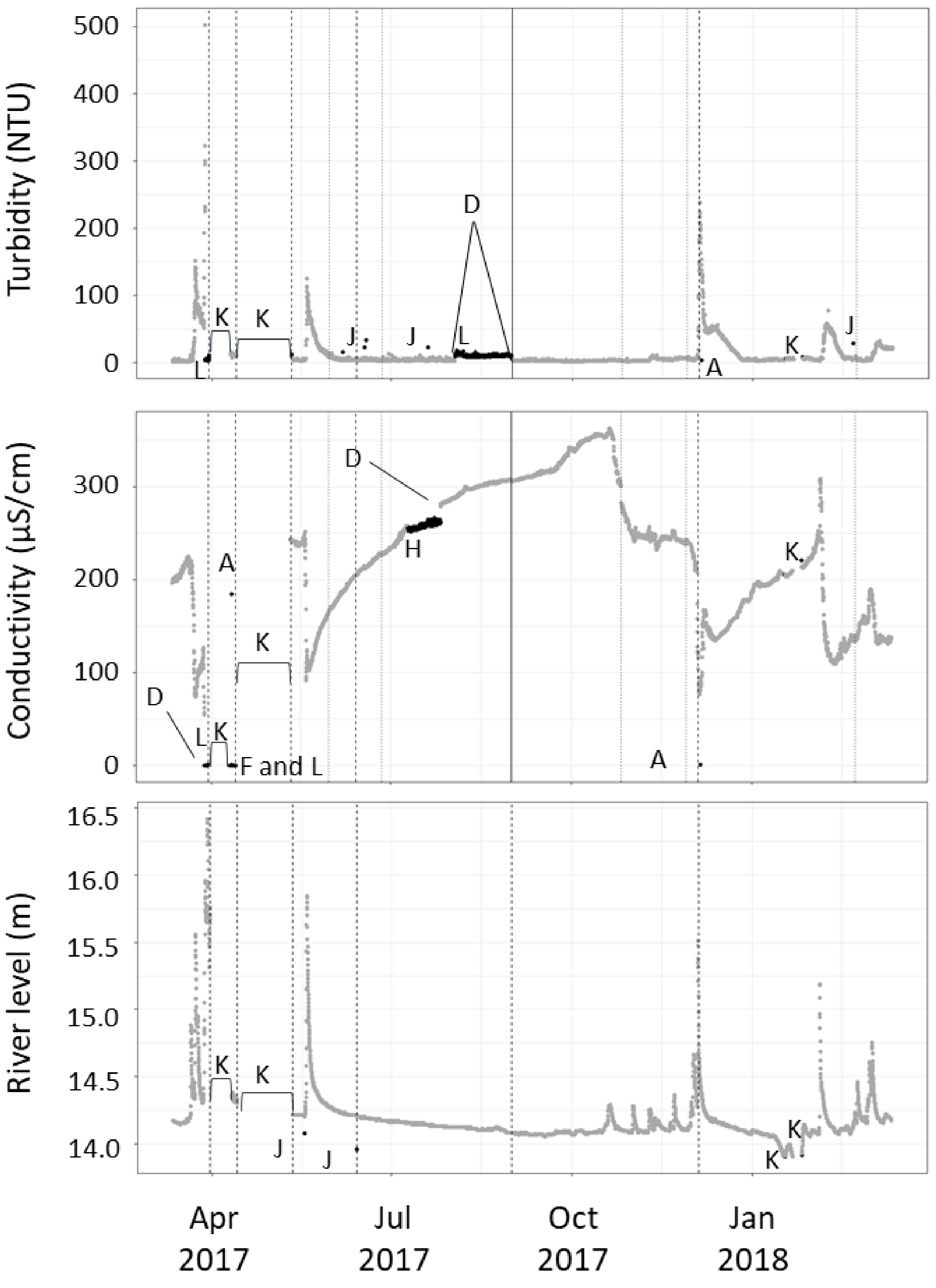

**Figure S2:** Time series for turbidity (NTU), conductivity (µS/cm) and river level (m) measured by *in situ* sensors at Pioneer River (PR). Dark grey points show non-anomalous data and black points show potentially anomalous data, identified by the end-user and labelled by type as per Table 1 in the main article. Pale grey dotted vertical lines indicate times of calibrated probe swapping, dark grey dotted vertical lines indicate times of other maintenance activities, such as battery checks, solid dark grey vertical lines indicate times at which both types of activity occurred.

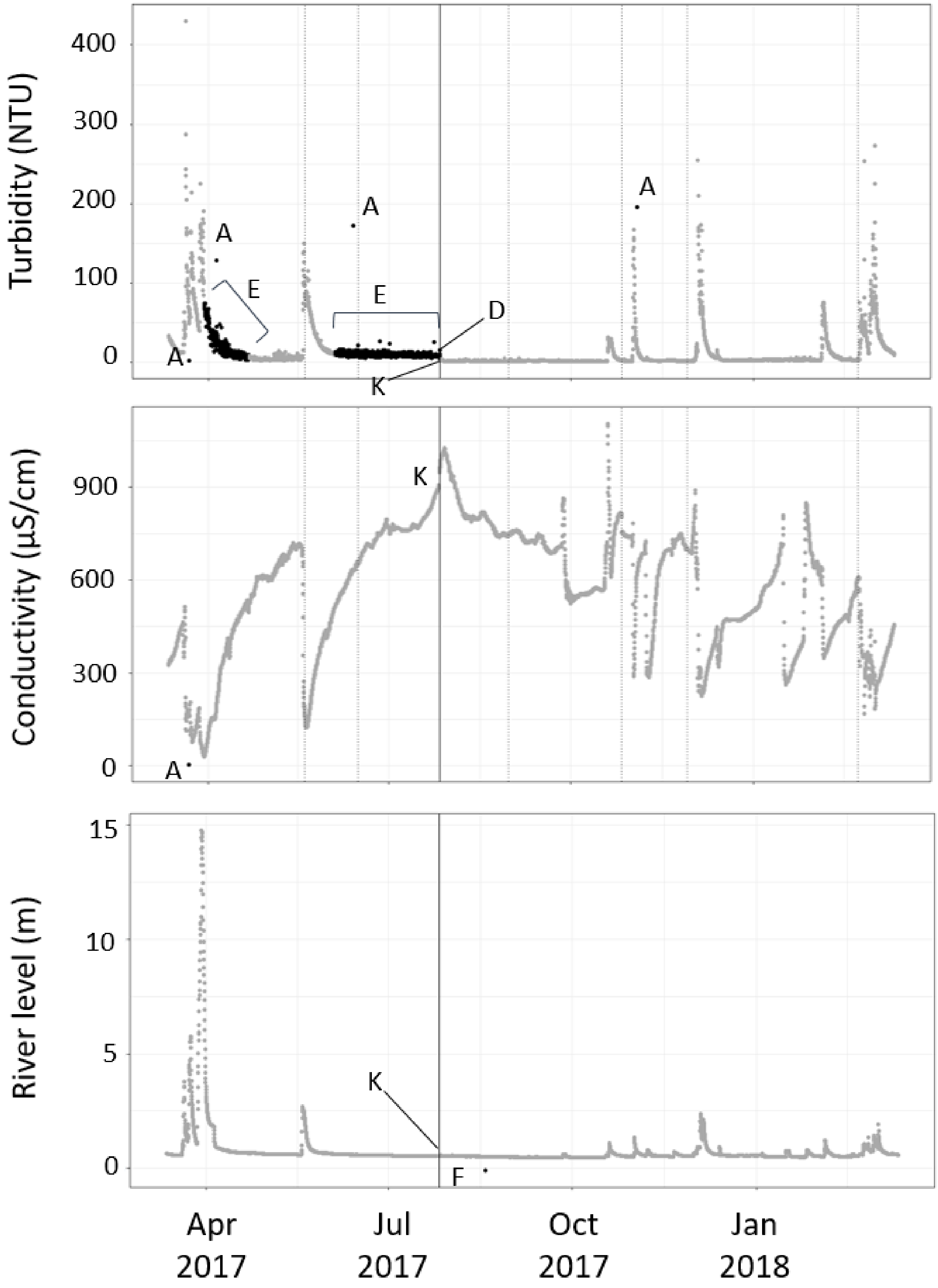

**Figure S3:** Time series for turbidity (NTU), conductivity (µS/cm) and river level (m) measured by *in situ* sensors at Sandy Creek (SC). Dark grey points show non-anomalous data and black points show potentially anomalous data, identified by the end-user and labelled by type as per Table 1 in the main article. Pale grey dotted vertical lines indicate times of calibrated probe swapping, dark grey dotted vertical lines indicate times of other maintenance activities, such as battery checks, solid dark grey vertical lines indicate times at which both types of activity occurred.

**Table S1:** Characteristics of anomaly detection methods used in the literature for water quality and/or time series data.

| Based on | Method | Learning method | Output | Dimensions | Probabilistic background | Near-real time data tractable | Big-data tractable | Nonstationary data tractable | Robust to missing values |
|---|---|---|---|---|---|---|---|---|---|
| Rules | Classification rules*[1] | n/a | C | U,M,MU | n | y | y | y | y |
| Regression | Naïve prediction*[2] | n/a | R | U | n | y | y | y | n |
| | Linear autoregression*[2] | SS | R | U | n | y | y | y | n |
| | Autoregressive integrated moving average (ARIMA)*[2] | SS | R | U | n | y | y | y | n |
| | Nearest cluster[2] | SS | R | U | n | y | y | y | n |
| | Multiple-level perceptron[2] | SS | R | U | n | y | y | y | n |
| Uncertainty fusion | GPR Uncertainty Fusion[3] | SS | R | U | y | y | - | y | - |
| Features | Box-modelling[4] | SS | C | MU | n | y | - | y | n |
| | OddStream[5] | SS | C | MU | y | y | y | y | - |
| | HDoutliers*[6] | U | C | U,M | y | n | y | y | - |
| | *k*-nearest neighbour (aggregated) (*k*NN-agg)*[7,8] | U | C | U,M | y | n | y | y | - |
| | *k*-nearest neighbour (summed) *k*NN-sum)*[8] | U | C | U,M | y | n | y | y | - |
| Dynamic Bayesian networks | Bayesian credible interval-Kalman filtering-uncoupled[9,10] | SS | R | U,MU,M | y | y | y | y | y |
| | Bayesian credible interval-robust Kalman filtering-uncoupled[9,10] | SS | RC | U,M,MU | y | y | y | y | y |
| | Maximum-A-Posteriori-uncoupled[9,10] | SS | RC | U,M,MU | y | y | y | y | y |
| Hidden Markov models | Fuzzy C-Means clustering[11] | SS | RC | U,M,MU | y | y | y | n | - |
| | Fuzzy integrals[11] | SS | RC | U,M,MU | y | y | y | n | - |
| Physical processes | Physical process models[12] | n/a | R | U,M,MU | n | n | - | y | y |

Notes: ‘Learning method’ indicates if the method requires labelling and/or training (S, supervised methods require a fully labelled training dataset, and test dataset; SS, semi-supervised methods require an anomaly-free training dataset, and test dataset; U, unsupervised methods do not require nor distinguish between training and test data); ‘Output’ indicates if the method does regression (R) and then anomaly classification, classification only (C), or or uses a mixture of both regression and classification (RC); ‘Dimensions’ refers to univariate (U), multivariate (multiple variables) (M), or multiple univariate (multiple streams of the same variable; MU) applications; ‘Probabilistic background’ indicates if the method uses probability theory to give an uncertainty level of the classification predictions; ‘Data tractable’ indicates if the method can deal with near-real time data, big data, or could learn if applied to nonstationary data; ‘Robust to missing values’ indicates if the method could still make classification predictions in the case of missing observations. Examples and/or as applied in [1]Fiebrich et al. (2010), [2]Hill and Minsker (2010), [3]Pang et al. (2016), [4]Chan and Mahoney (2005), [5]Talagala et al. (2018), [6]Wilkinson (2018), [7]Angiulli and Pizzuti (2002), [8]Madsen (2018), [9]Dereszynski and Dietterich (2007), [10]Hill et al. (2009), [11]Li et al. (2017), and [12]Moatar et al. (2001). n/a, not applicable; n, no; y, yes; -, information not provided in the relevant paper. *Methods implemented in this study.

**Table S2:** Confusion matrix based on binary classification of observations as anomalies or not.

| | **Actual anomaly (positive class)** | **Actual non-anomaly (negative class)** |
|---|---|---|
| Classified anomaly (positive class) | True positive (TP) | False positive (FP) |
| Classified non-anomaly (negative class) | False negative (FN) | True Negative (TN) |

**Table S3:** Metrics used in this study to compare and assess the performance of different anomaly detection methods.

| Based on: | Metric | Formula | Defines the: | Interpretation | Advantages / Disadvantages |
|---|---|---|---|---|---|
| Confusion matrix | Accuracy[1,2] | (TP+TN) / (TP + FP + TN + FN) | Proportion of all observations correctly classified. | Higher values are more preferable. | Easy to calculate and interpret/ Does not capture poor performance of classification for unbalanced datasets (where normal observations > anomalous observations or vice versa) |
| | Error rate[1,2] | (FP + FN ) / (TP + FP + TN + FN) | Proportion of all observations incorrectly classified. Higher values are more preferable | Lower values are more preferable. | As above |
| | PPV[3] | TP / (TP + FP) | Proportion of classified anomalies correctly identified | Higher values are more preferable, especially when NPV is also high. | Good for unbalanced datasets where normal observations > anomalous observations and when sensitivity is important (classification of anomalies as anomalies)/ - |
| | NPV[3] | TN / (TN + FN) | Proportion of classified non-anomalous observations correctly identified | Higher values are more preferable, especially when PPV is also high. | Good for unbalanced datasets where normal observations > anomalous observations and when specificity is important (classification of non-anomalous data as normal)/ - |
| Regression model | RMSE[1] | $\sqrt{\frac{\sum_{i=1}^{n}(y_i - \hat{y}_i)^2}{n}}$ | Difference between the predicted solutions and desired solutions | Lower values are more preferable. | Commonly used and easy to calculate / Not applicable for all anomaly detection methods (e.g. does not apply to the feature-based methods used herein) |

Notes: [1]Hossin and Sulaiman (2015), [2]Sokolova and Lapalme (2009), [3]Ranawana and Palade (2006). TP, true positive; TN, true negative; FP, false positive; FN, false negative (Table 4); PPV and NPV, positive and negative predictive values, respectively; RSME, root mean square error.

**Table S4:** Performance metrics for regression-based methods of anomaly detection performed on conductivity data from an in situ sensor at Sandy Creek (SC), incorporating 100% detection of Class 2 anomalies by automated classification rules. See Tables S2-3 for metric formula and descriptions. Numbers in parenthesis indicate *p*, *d* and *q* parameters as per the ARIMA form of each model.

| Model | Method | TN | FN | FP | TP | Accuracy | Error rate | NPV | PPV | RMSE |
|---|---|---|---|---|---|---|---|---|---|---|
| Naïve (0,1,0) | AD | 5340 | 0 | 60 | 2 | 0.99 | 0.01 | 1.00 | 0.03 | 0.08 |
| | ADAM | 859 | 0 | 4541 | 2 | 0.16 | 0.84 | 1.00 | 0.00 | 0.08 |
| Linear AR (3,0,0) | AD | 5322 | 0 | 78 | 2 | 0.99 | 0.01 | 1.00 | 0.03 | 0.09 |
| | ADAM | 3988 | 0 | 1412 | 2 | 0.74 | 0.26 | 1.00 | 0.00 | 0.45 |
| ARIMA (2,1,3) | AD | 5361 | 0 | 39 | 2 | 0.99 | 0.01 | 1.00 | 0.05 | 0.08 |
| | ADAM | 3994 | 0 | 1406 | 2 | 0.74 | 0.26 | 1.00 | 0.00 | 0.45 |
| RegARIMA (3,1,0) | AD | 5284 | 0 | 116 | 2 | 0.98 | 0.02 | 1.00 | 0.02 | 0.13 |
| | ADAM | 309 | 0 | 5091 | 2 | 0.06 | 0.94 | 1.00 | 0.00 | 0.08 |

Notes: AD, anomaly detection; ADAM, anomaly detection and mitigation; AR, autoregression.

**Table S5:** Performance metrics for feature-based methods of anomaly detection performed on multivariate time series data from *in situ* sensors at Sandy Creek (SC), for conductivity, incorporating 100% detection of Class 2 anomalies by automated classification rule. See Tables S2-3 for metric formula and descriptions.

| Method | Transformation | TN | FN | FP | TP | Accuracy | Error rate | NPV | PPV |
|---|---|---|---|---|---|---|---|---|---|
| HDoutliers | Derivative | 5398 | 1 | 2 | 1 | 1.00 | 0.00 | 0.33 | 1.00 |
| | OS Derivative | 5399 | 1 | 1 | 1 | 1.00 | 0.00 | 0.50 | 1.00 |
| *k*NN_agg | Derivative | 5395 | 1 | 5 | 1 | 1.00 | 0.00 | 0.17 | 1.00 |
| | OS Derivative | 5367 | 1 | 33 | 1 | 0.99 | 0.01 | 0.03 | 1.00 |
| *k*NN_sum | Derivative | 5396 | 1 | 4 | 1 | 1.00 | 0.00 | 0.20 | 1.00 |
| | OS Derivative | 5367 | 1 | 33 | 1 | 0.99 | 0.01 | 0.03 | 1.00 |

Notes: OS, one-sided.

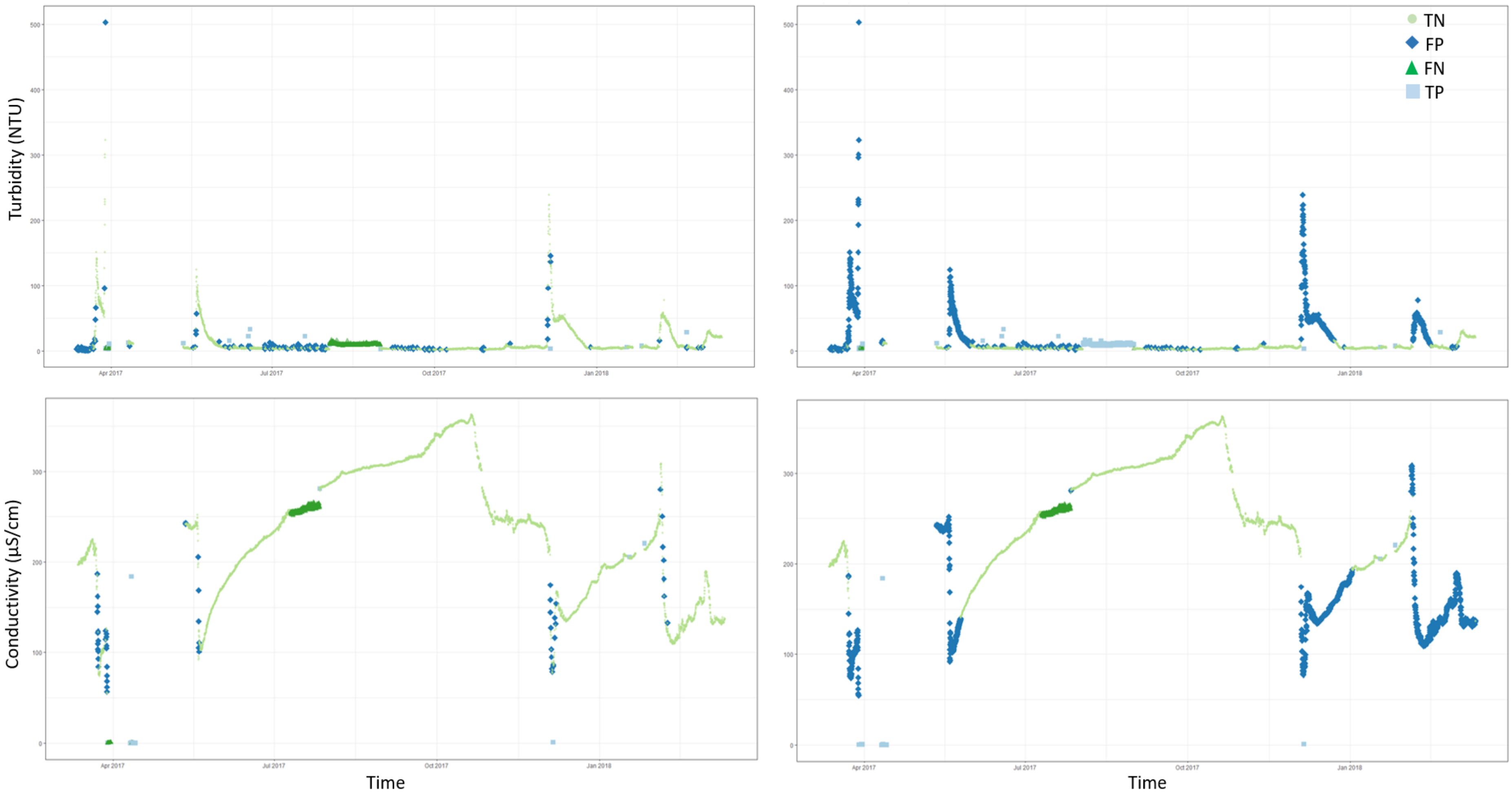

**Figure S4:** Classification of turbidity (upper row) and conductivity observations (lower row) measured by *in situ* sensors at Pioneer River (PR) by linear autoregression as true negatives (TN), false negatives (FN), false positives (FP) or true positives (TP). Plots on the left show results from linear autoregression alone, those on the right show results from linear autoregression with ADAM.

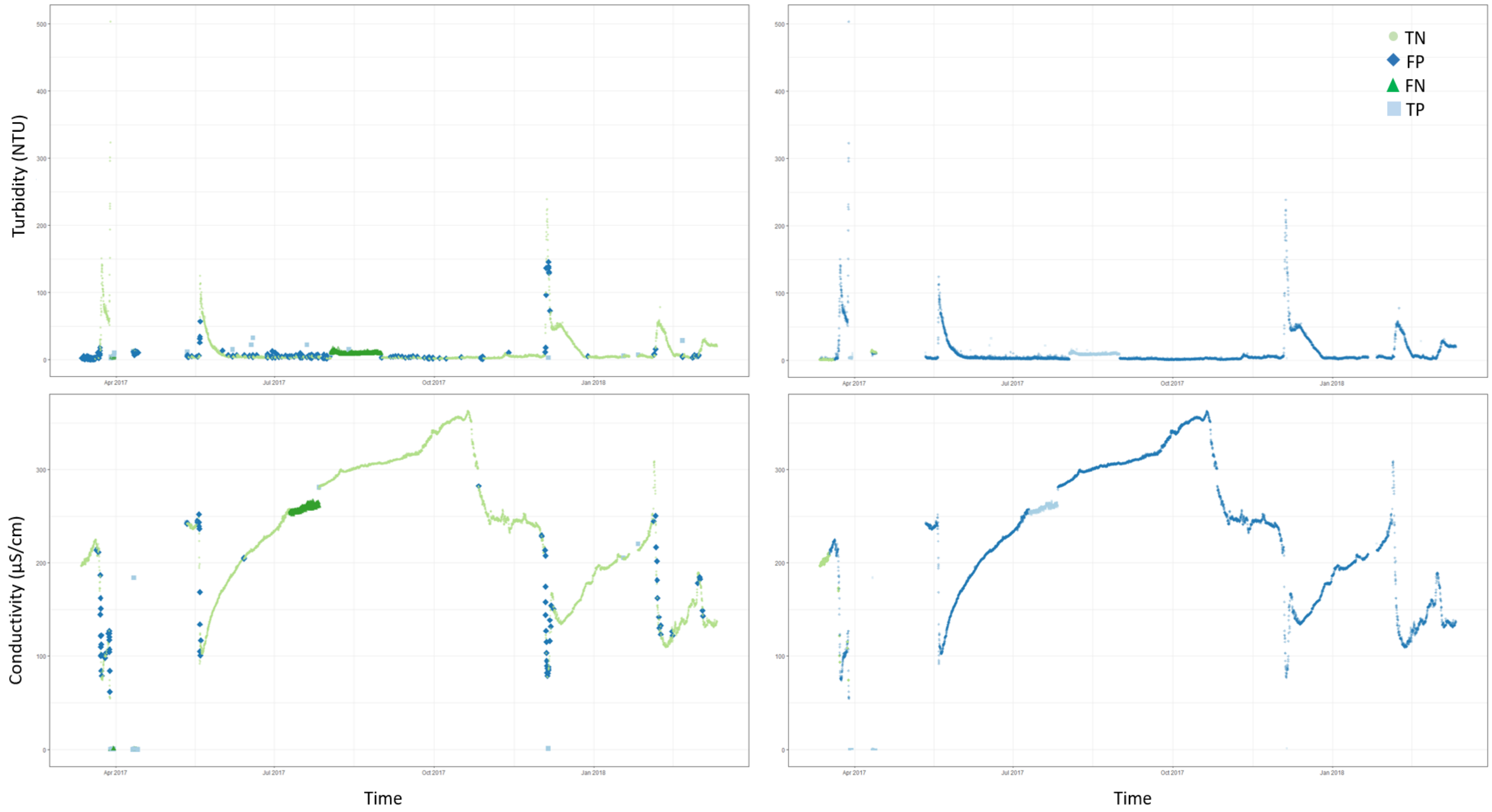

**Figure S5:** Classification of turbidity (upper row) and conductivity observations (lower row) measured by *in situ* sensors at Pioneer River (PR) by RegARIMA as true negatives (TN), false negatives (FN), false positives (FP) or true positives (TP). Plots on the left show results from RegARIMA alone, those on the right show results from RegARIMA with ADAM.

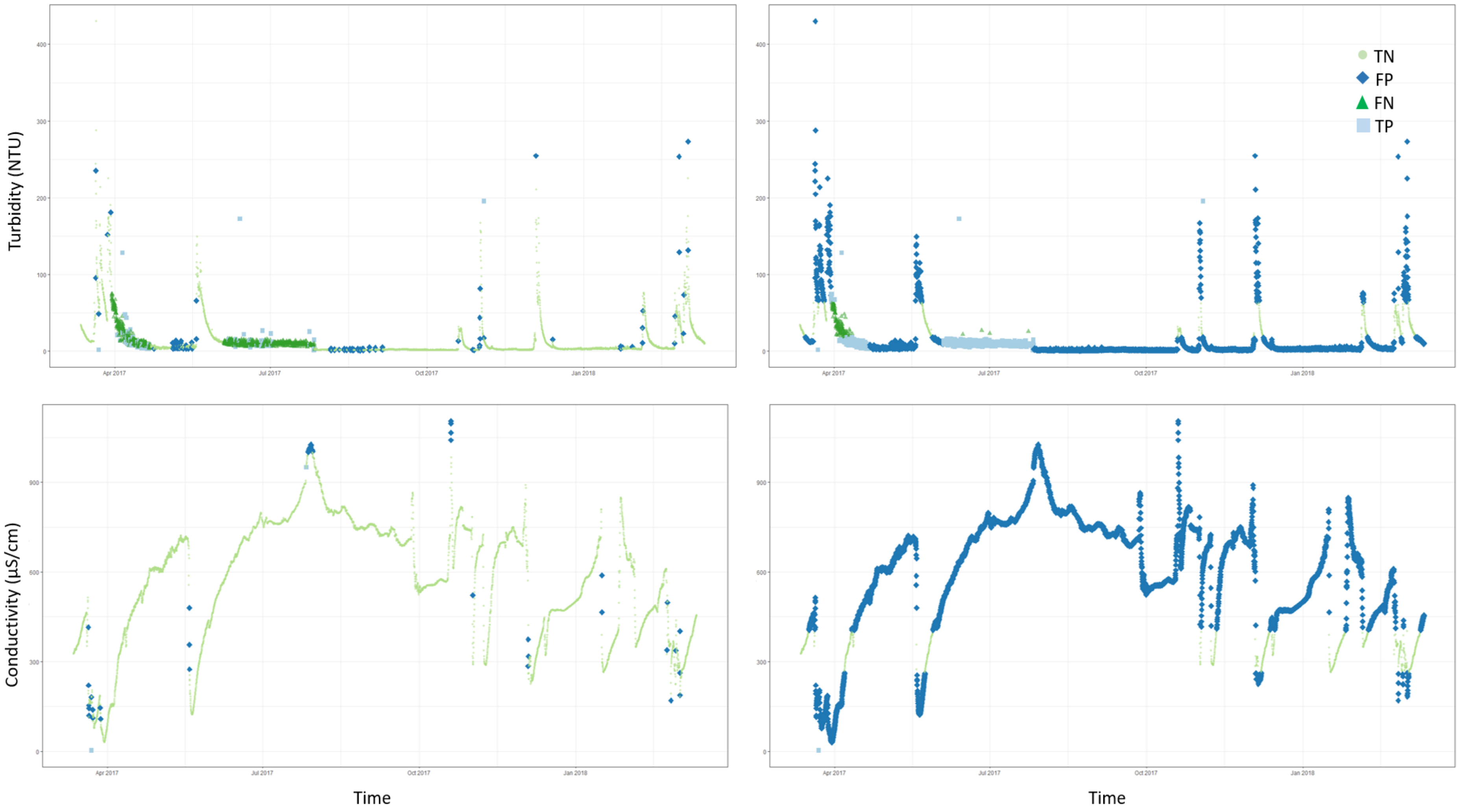

**Figure S6:** Classification of turbidity (upper row) and conductivity observations (lower row) measured by *in situ* sensors at Sandy Creek (SC) by naïve prediction as true negatives (TN), false negatives (FN), false positives (FP) or true positives (TP). Plots on the left show results from naïve prediction alone, those on the right show results from naïve prediction with ADAM.

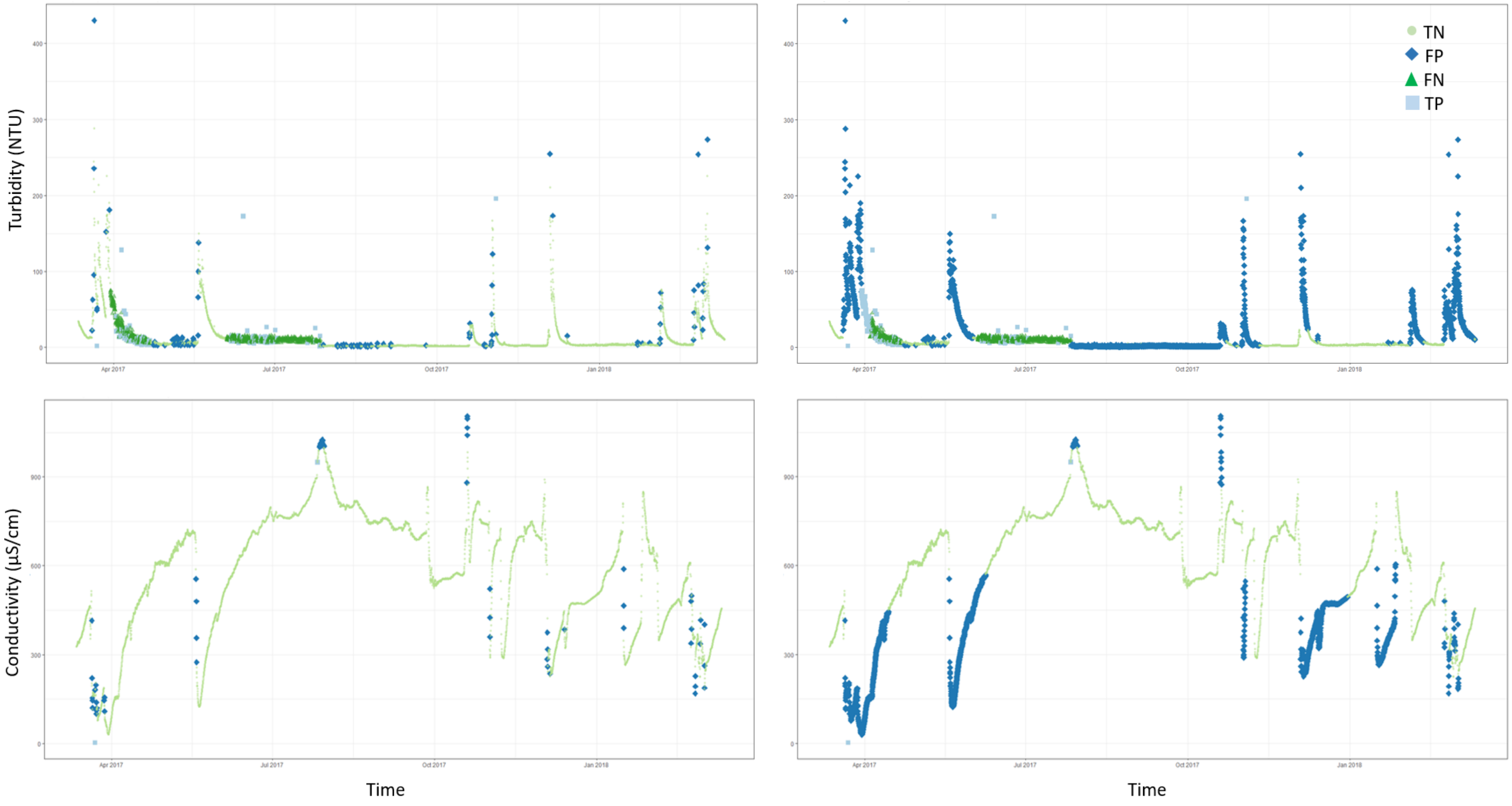

**Figure S7:** Classification of turbidity (upper row) and conductivity observations (lower row) measured by *in situ* sensors at Sandy Creek (SC) by linear autoregression as true negatives (TN), false negatives (FN), false positives (FP) or true positives (TP). Plots on the left show results from linear autoregression alone, those on the right show results from linear autoregression with ADAM.

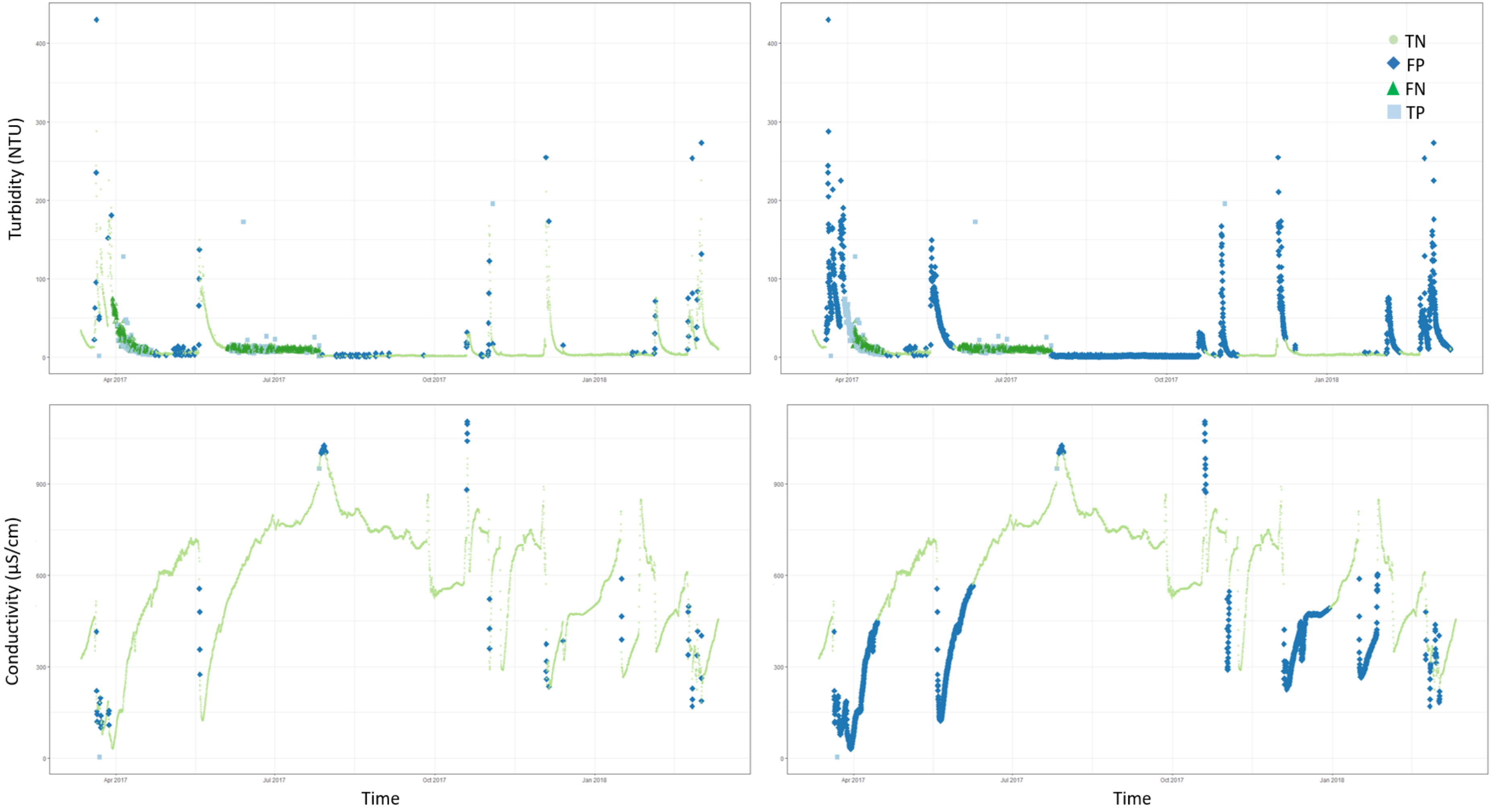

**Figure S8:** Classification of turbidity (upper row) and conductivity observations (lower row) measured by *in situ* sensors at Sandy Creek (SC) by ARIMA as true negatives (TN), false negatives (FN), false positives (FP) or true positives (TP). Plots on the left show results from ARIMA alone, those on the right show results from ARIMA with ADAM.

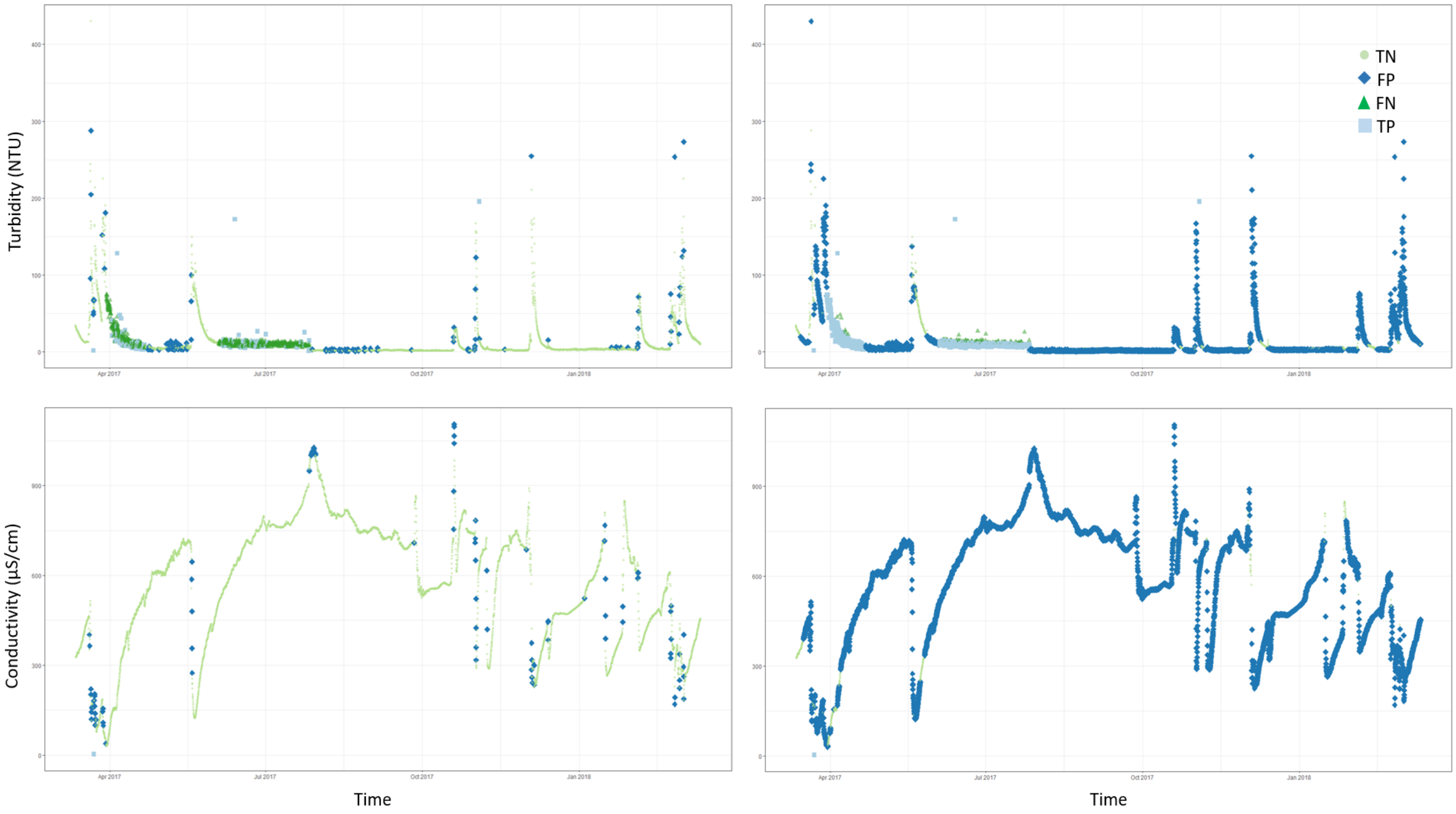

**Figure S9:** Classification of turbidity (upper row) and conductivity observations (lower row) measured by *in situ* sensors at Sandy Creek (SC) by RegARIMA as true negatives (TN), false negatives (FN), false positives (FP) or true positives (TP). Plots on the left show results from RegARIMA alone, those on the right show results from RegARIMA with ADAM.

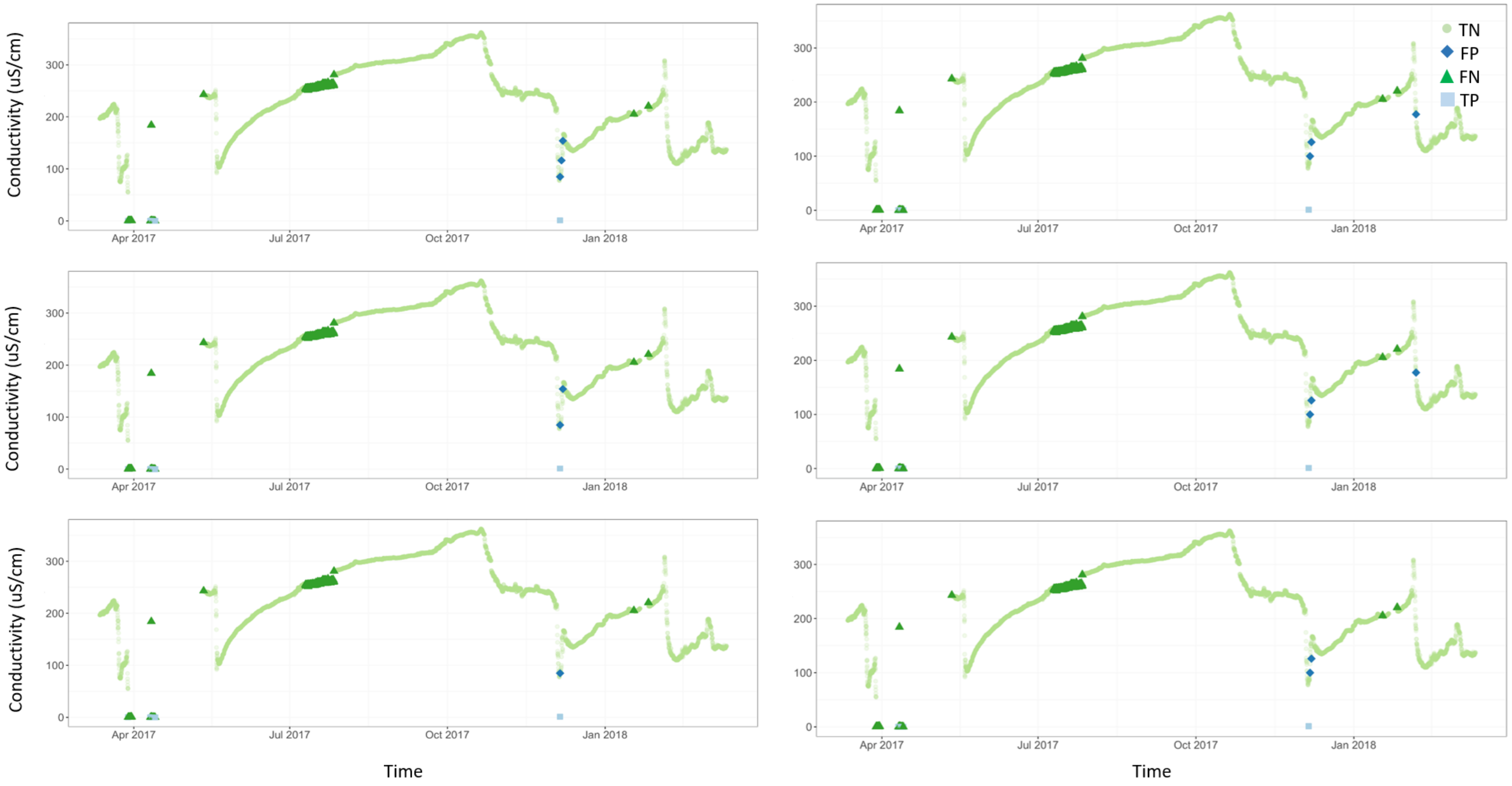

**Figure S10:** Classification of conductivity measured by an *in situ* sensor at Pioneer River (PR) by HDoutliers (upper row), *k*NN-agg (middle row) and *k*NN-sum (lower row) as true negatives (TN), false negatives (FN), false positives (FP) or true positives (TP). Plots on the left show results of methods applied to the derivatives, and those on the right show results of methods applied to the one-sided derivatives of the time series.

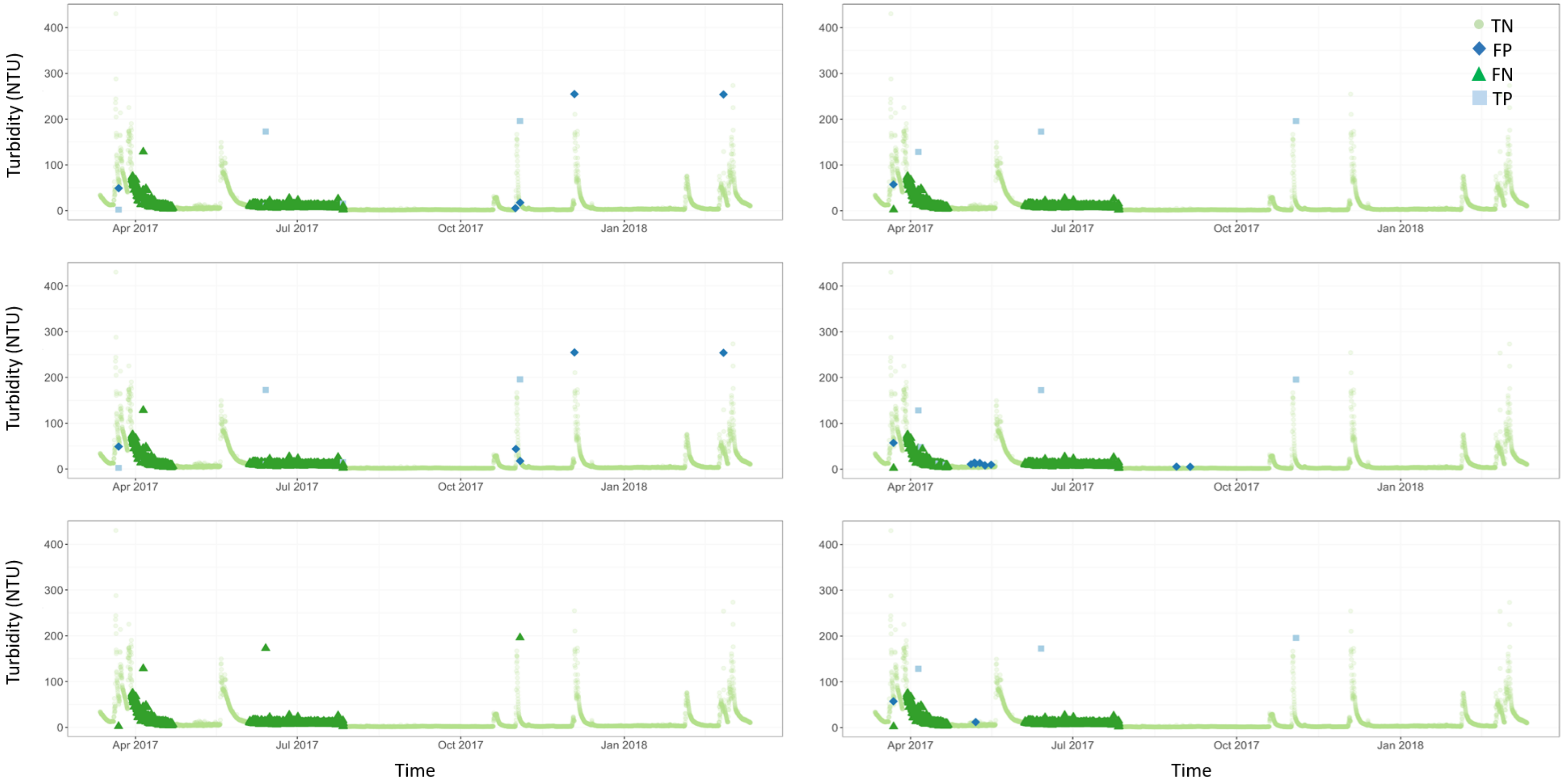

**Figure S11:** Classification of turbidity measured by an *in situ* sensor at Sandy Creek (SC) by HDoutliers (upper row), *k*NN-agg (middle row) and *k*NN-sum (lower row) as true negatives (TN), false negatives (FN), false positives (FP) or true positives (TP). Plots on the left show results of methods applied to the derivatives, and those on the right show results of methods applied to the one-sided derivatives of the time series.

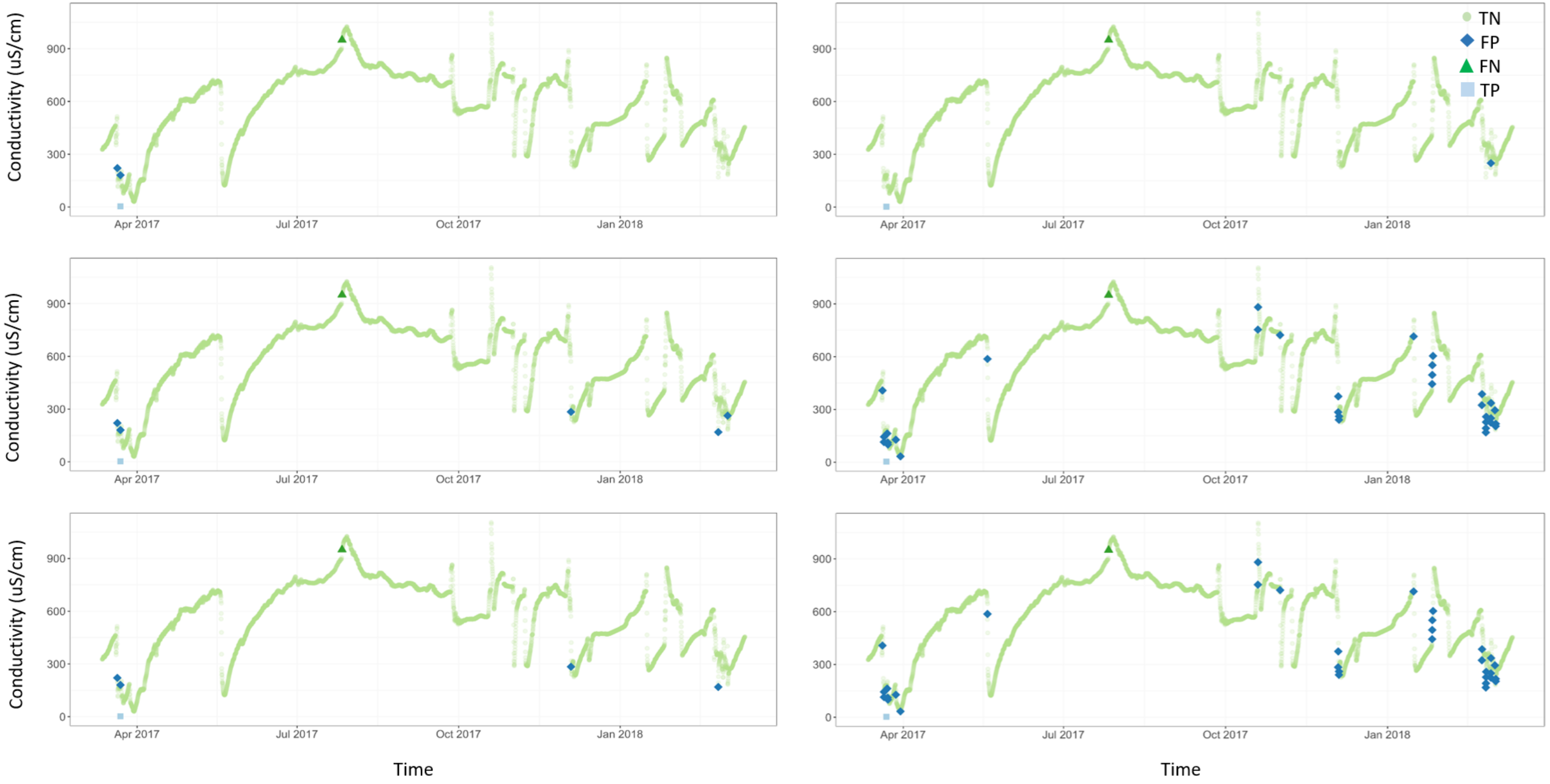

**Figure S12:** Classification of conductivity measured by an *in situ* sensor at Sandy Creek (SC) by HDoutliers (upper row), *k*NN-agg (middle row) and *k*NN-sum (lower row) as true negatives (TN), false negatives (FN), false positives (FP) or true positives (TP). Plots on the left show results of methods applied to the derivatives, and those on the right show results of methods applied to the one-sided derivatives of the time series.

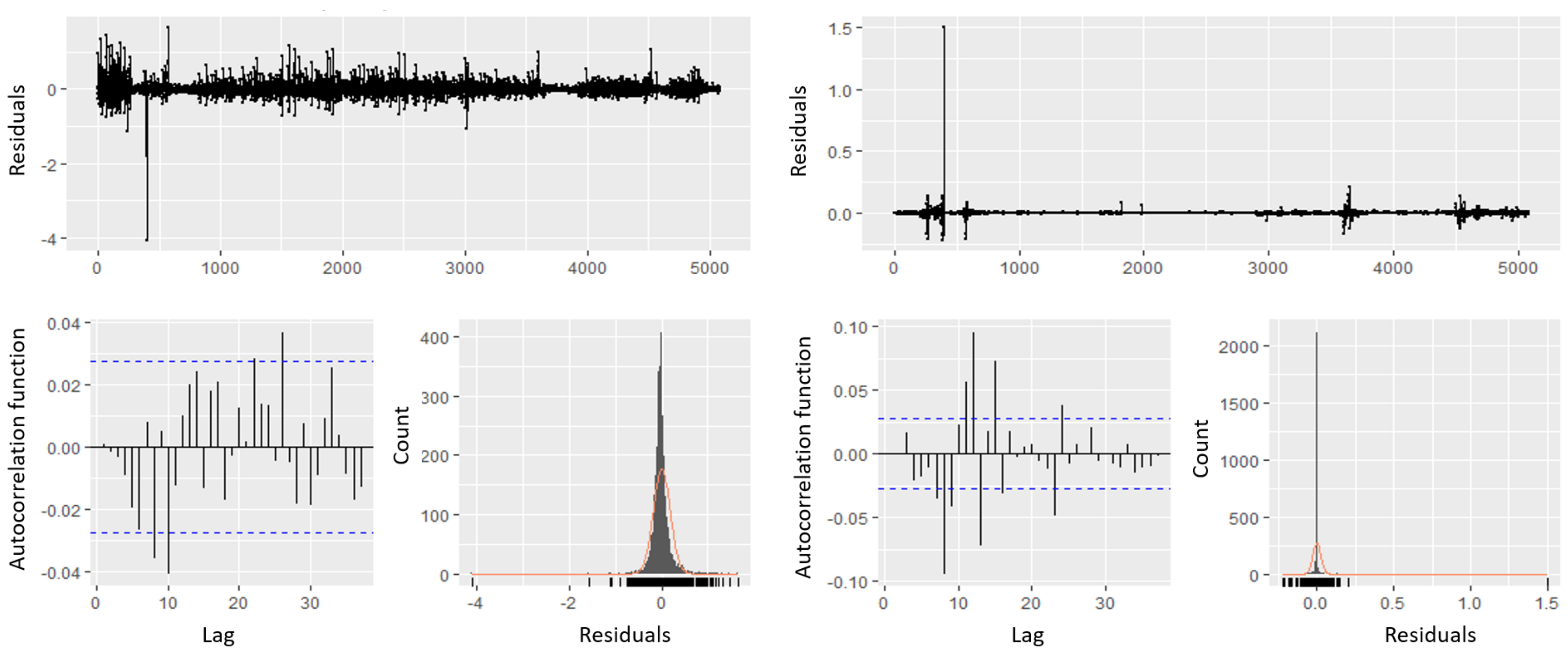

**Figure S13:** Diagnostic plots from the linear autoregression applied to the turbidity (left-most three plots) and conductivity time series (right-most three plots) at Pioneer River (PR).

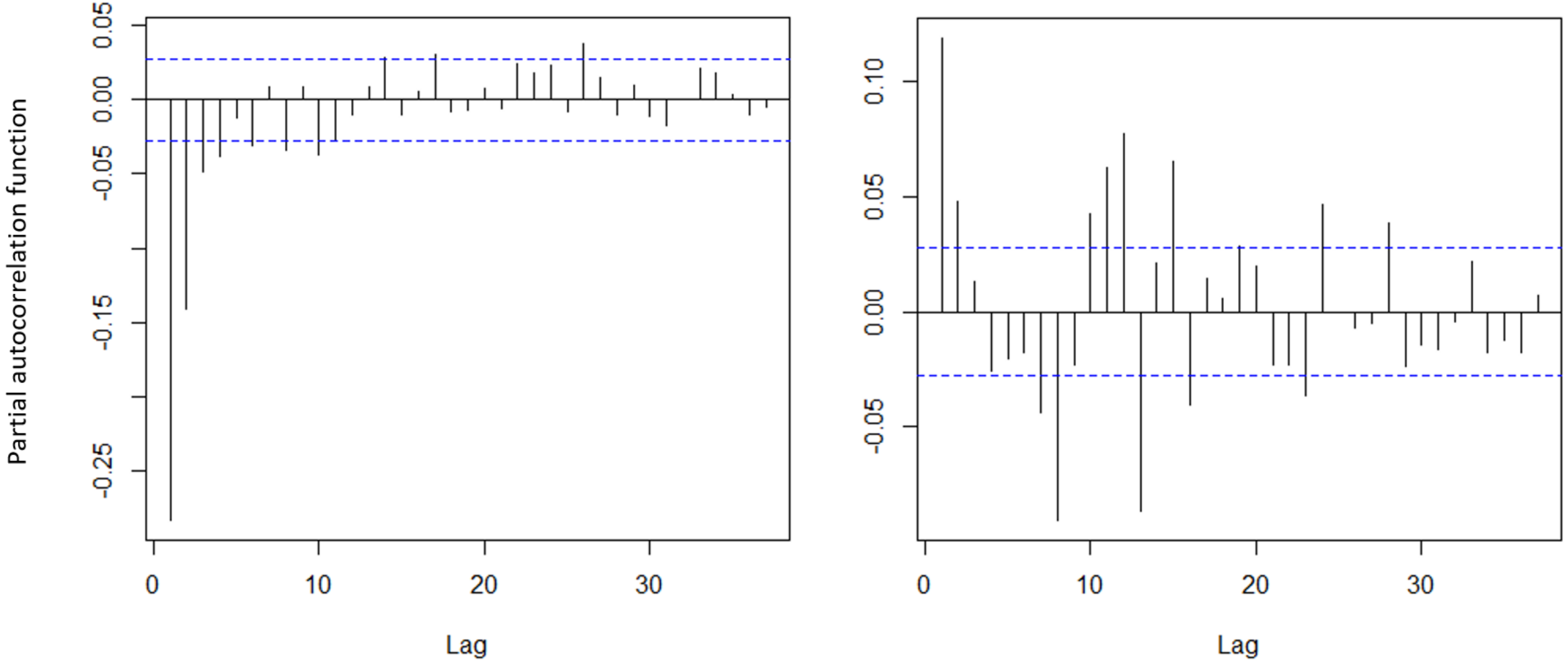

**Figure S14:** Partial autocorrelation function (PACF) plots from the linear autoregression for the turbidity (left) and conductivity time series (right) at Pioneer River (PR).

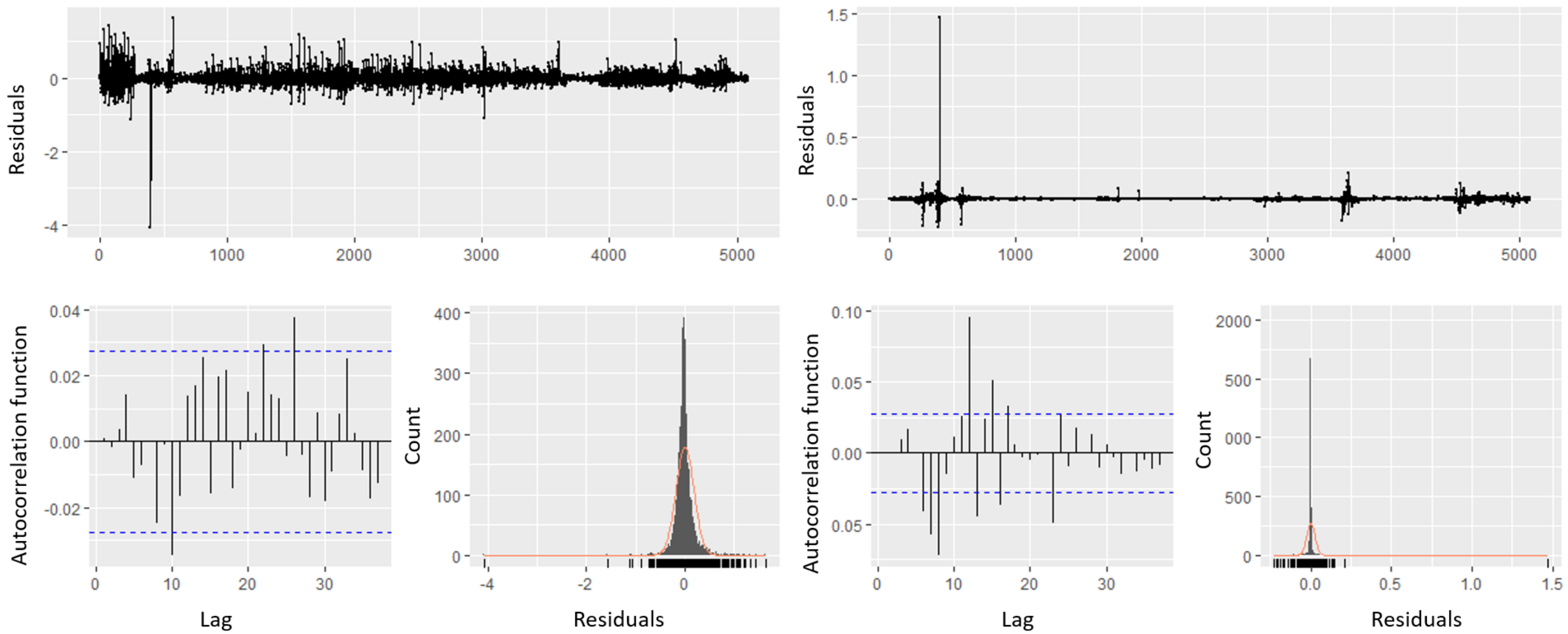

**Figure S15:** Diagnostic plots from the ARIMA applied to the turbidity (left-most three plots) and conductivity time series (right-most three plots) at Pioneer River (PR).

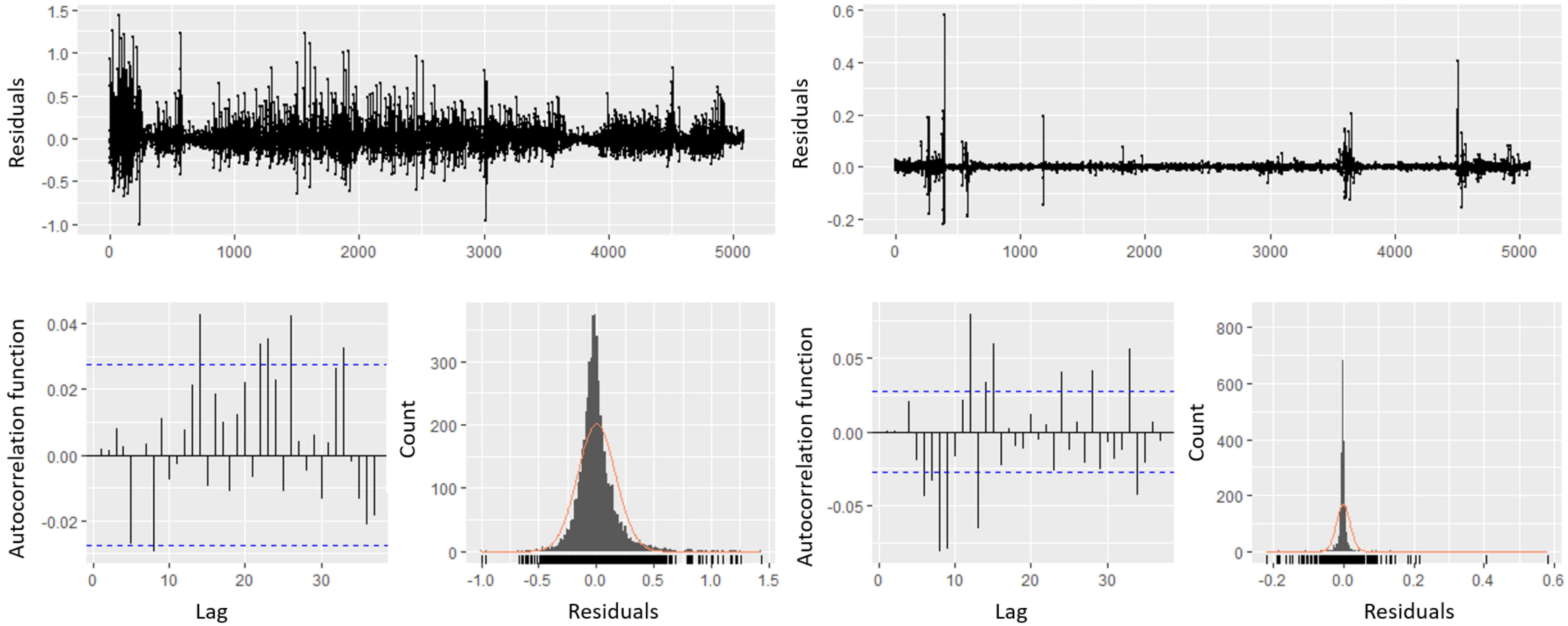

**Figure S16:** Diagnostic plots from the RegARIMA applied to the turbidity (left-most three plots) and conductivity time series (right-most three plots) at Pioneer River (PR).

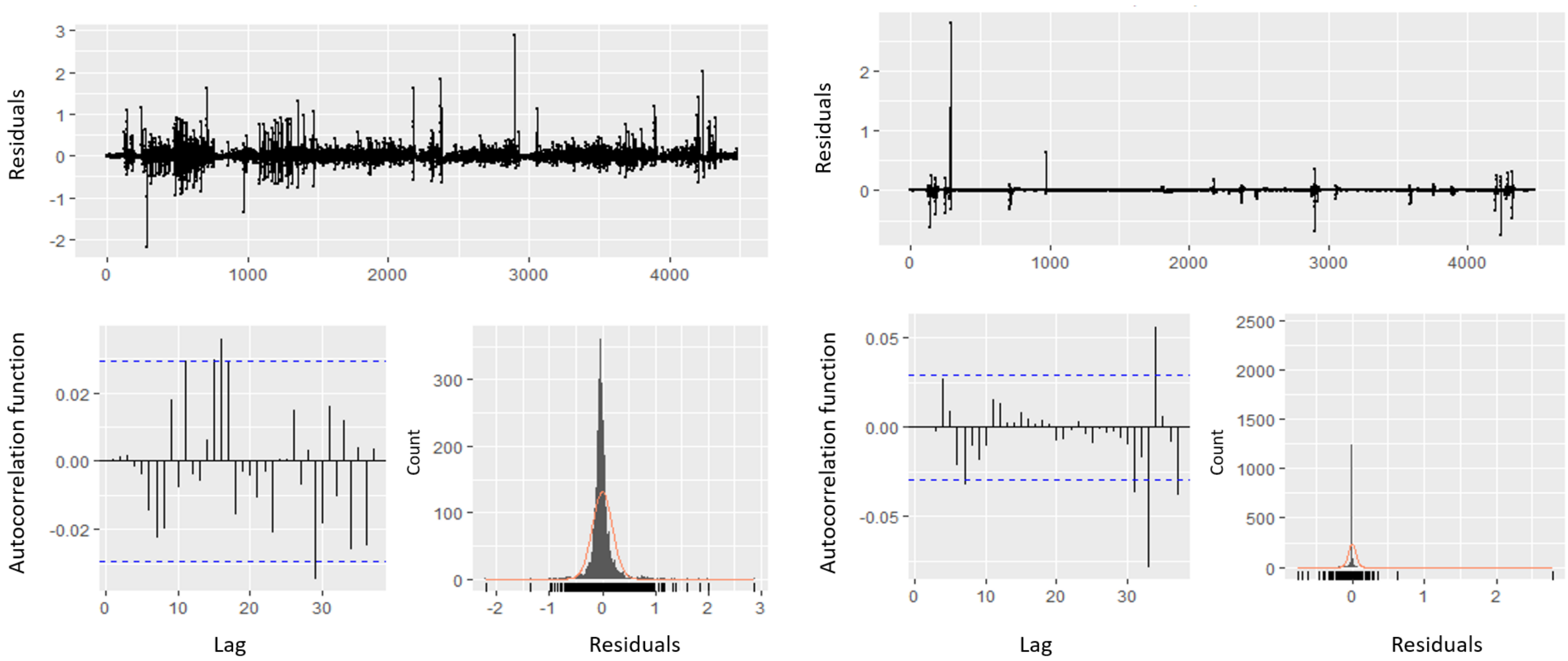

**Figure S17:** Diagnostic plots from the linear autoregression applied to the turbidity (left-most three plots) and conductivity time series (right-most three plots) at Sandy Creek (SC).

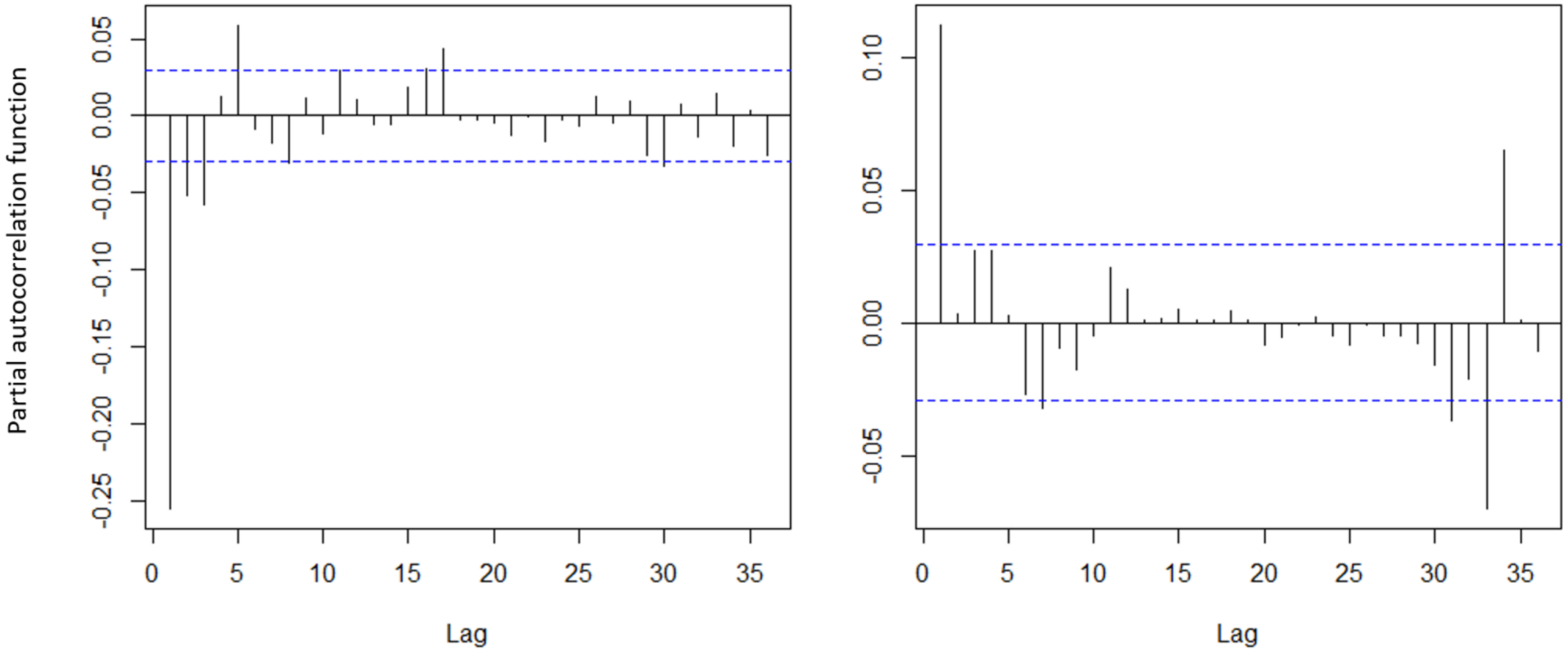

**Figure S18:** Partial autocorrelation function (PACF) plots from the linear autoregression for the turbidity (left) and conductivity time series (right) at Sandy Creek (SC).

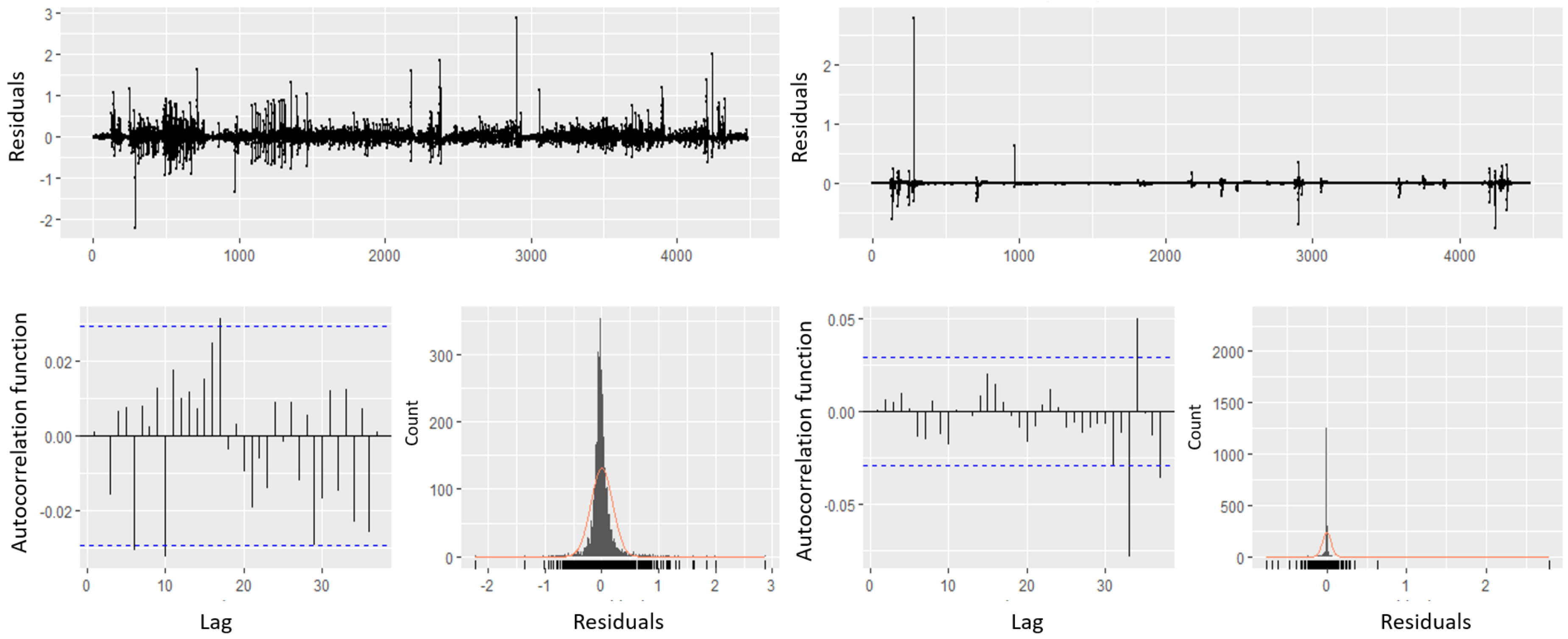

**Figure S19:** Diagnostic plots from the ARIMA applied to the turbidity (left-most three plots) and conductivity time series (right-most three plots) at Sandy Creek (SC).

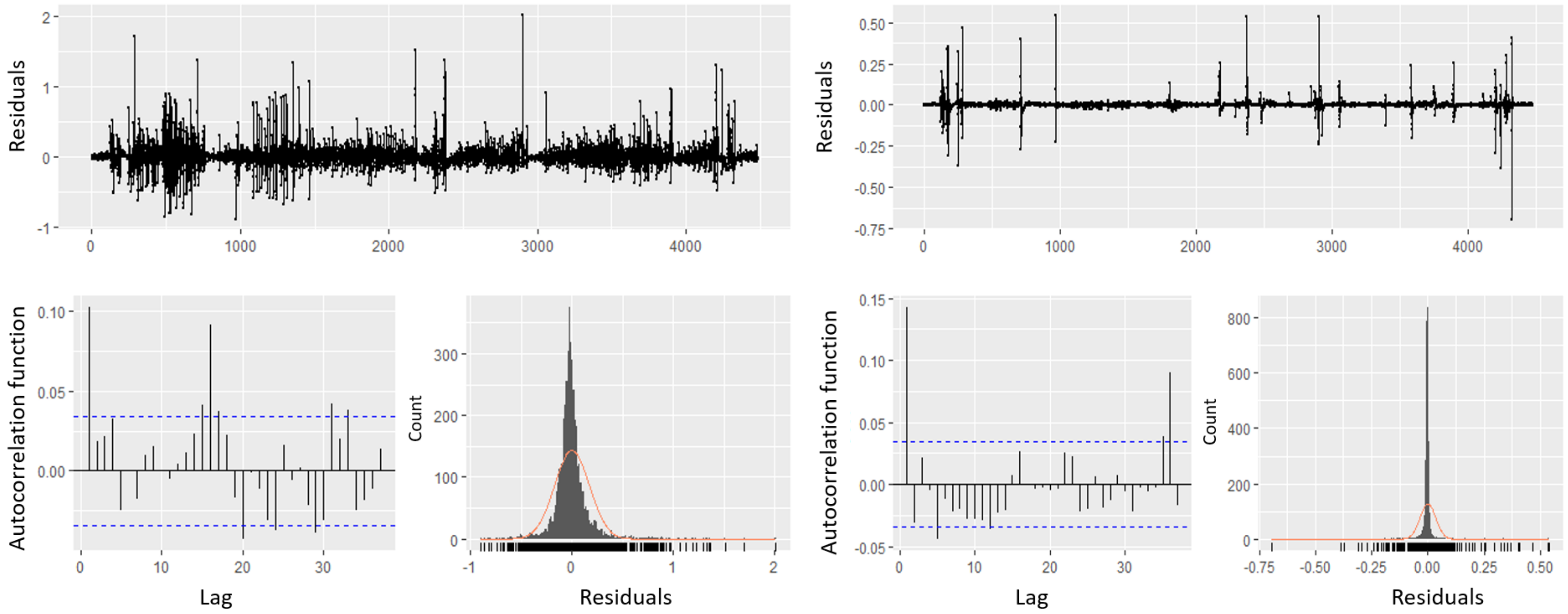

**Figure S20:** Diagnostic plots from the RegARIMA applied to the turbidity (left-most three plots) and conductivity time series (right-most three plots) at Sandy Creek (SC).